\documentclass[twocol]{ametsocV6.1}
\usepackage{comment} 

\title{On the Links Between Thermobaricity, Available Potential Energy, Neutral Directions, Buoyancy Forces, and Lateral Stirring in the Ocean}

\authors{R\'emi Tailleux \aff{a}\correspondingauthor{R. Tailleux, R.G.J.Tailleux@reading.ac.uk}
Gabriel Wolf, \aff{a}} 

\affiliation{\aff{a}{Department of Meteorology, University of Reading, Reading, United Kingdom}}

\abstract{\textcolor{black}{The various ingredients of Lorenz theory of available potential energy (APE) are shown to hold the key for understanding how to develop a first-principles theory of lateral stirring and lateral stirring surfaces in the oceans embedded in the study of the full Navier-Stokes equations for compressible seawater. This theory establishes that it is the existence of thermobaric forces acting along isopycnal surfaces that makes stirring in seawater fundamentally different from that in a simple fluid and is the ultimate cause for the non-existence of neutral surfaces. It also establishes that the `true' neutral directions are those perpendicular to an APE-based form of the P vector previously identified by Nycander, contrary to what has been assumed so far. Where thermobaric forces are small enough to be neglected, our theory establishes that the Lorenz reference density (LRD) surfaces entering APE theory are very accurately neutral and represent the relevant definition of lateral stirring surfaces. Where thermobaric forces are large, however, lateral stirring becomes strongly coupled with vertical stirring and complicates the identification of the `right' lateral mixing surfaces. Importantly, rewriting the momentum balance equations in their thermodynamic form using Crocco-Vazsonyi theorem and removing the dynamically inert part of the Bernoulli function proves decisive for obtaining most results. The new results have important implications for the theory of isopycnal analysis and ocean mixing parameterisations.} } 

\begin{document} 

\maketitle

%
%
%
\statement
\textcolor{black}{The theoretical justification for the `lateral mixing surfaces', a.k.a. isopycnal surfaces, along which the ocean water masses are thought to spread away from their formation regions to then mix with other water masses in the ocean interior has so far primarily relied on heuristic and ad-hoc considerations rooted in two-parcel arguments. Unfortunately, because such arguments do not lead to testable predictions, how to test their validity in observations and models has remained unclear so far. The physical significance of this work lies in that it represents the first successful attempt of its kind at rooting the theory of water masses and isopycnal surfaces in the study of the full governing equations of motion. This makes it capable of producing testable predictions, which are found to often depart from popular wisdom and call into question many aspects of how the problem has been approached so far with implications for how to parameterise turbulent mixing in numerical ocean models and how to construct isopycnal surfaces. }
%
%
%

%

\section{Introduction} 

\textcolor{black}{The isentropic surfaces $\theta = {\rm constant}$ of a simple stratified fluid (such as dry air or pure water), where $\theta$ is potential temperature, play a central role in the study of mixing and stirring. One of the main reasons is because the deformations of such surfaces, which measure the changes in available potential energy (APE) caused by vertical stirring, are linked to the APE dissipation rate $\varepsilon_p$ and the turbulent diapycnal mixing diffusivity $K_{\rho} = \varepsilon_p/N^2$, e.g.,  \citet{Oakey1982,Gargett1984b,Winters1995,Winters1996,Lindborg2008,Gregg2021}, where $N^2$ is a suitably defined mean squared buoyancy frequency. Conversely, the notional form of stirring that leave the isentropic surfaces unaffected naturally defines `lateral stirring'. Physically, the lateral directions, being those perpendicular to $\nabla \theta$, are obviously local in character. In contrast, vertical stirring and $\varepsilon_p$, which depend on the global ocean stratification, are non-local in character due to buoyancy forces being proportional to the distance to an equilibrium state of the fluid, e.g., \citet{Dewar2019}, \citet{Taylor2019} and references therein. This local versus non-local dichotomy of lateral versus vertical stirring indicates that these two fundamental forms of stirring are decoupled in simple fluids. A key result of this paper will be to show that this property no longer holds in seawater due to thermobaricity, in contrast to what has been implicitly assumed so far, and that this is what makes the identification of lateral stirring surfaces difficult in the oceans.}
\par 

\par
\textcolor{black}{Presumably because diapycnal mixing $\varepsilon_p$ can only exist if there is APE to be dissipated, oceanographers originally hypothesised that the lateral stirring directions in seawater should be those along which stirring minimally affect the distribution of mass and potential energy of the oceans. Using a two-parcel view of stirring, with $(S_1,\theta_1,p_1)$ and $(S_2,\theta_2,p_2)$ denoting the thermodynamic properties of the two parcels, \citet{Sverdrup1942} established that for the distribution of mass to remain unaffected following their interchange, the parcels' densities would need to satisfy} 
\begin{equation}
\begin{split} 
      \rho(S_1,\theta_1,p_2) & = \rho(S_2,\theta_2,p_2) ,  \\  \rho(S_2,\theta_2,p_1) & = \rho(S_1,\theta_1,p_1) ,
      \label{sverdrup_relations} 
\end{split} 
\end{equation}  
\textcolor{black}{However, because (\ref{sverdrup_relations}) has no exact solution in general, \citet{Sverdrup1942} proposed} that $\sigma_t = \rho(S,T,p_a) - 1000$, which had formed the basis for early isopycnal analyses \citep{Montgomery1938,Iselin1939}, be regarded as the \textcolor{black}{next} best practical alternative for defining lateral stirring directions, $p_a$ being the mean surface atmospheric pressure.

\par

\textcolor{black}{Next, oceanographers tried to turn Eq. (\ref{sverdrup_relations})} into a more tractable problem by only requiring the two parcel densities to equate at the mid-pressure $\overline{p} = (p_1+p_2)/2$ instead, viz. 
\begin{equation}
    \rho(S_1,\theta_1,\overline{p}) = \rho(S_2,\theta_2,\overline{p}) ,
    \label{neutral_path} 
\end{equation} 
which seemingly describes the parcel interchange as taking place on a {\em locally-referenced potential density} surface $\rho^{\ell} = \rho(S,\theta,\overline{p}) = {\rm constant}$. \textcolor{black}{Eq. (\ref{neutral_path}) has played a central role in the development of isopycnal analysis, as it underlies the `neutrality property' that forms the basis for \citet{Jackett1997}'s construction of $\gamma^n$, while being also very close to the condition used by \citet{Foster1976} to define their `lateral mixing paths'. Next, oceanographers considered replacing the parcel-dependent pressure $\overline{p}$ in (\ref{neutral_path}) by a fixed reference pressure $p_r$ representative of local conditions, which led to the well known and} widely used concept of potential density $\sigma_r = \rho(S,\theta,p_r)-1000$, but as is well known, the approximation deteriorates as $|\overline{p}-p_r|$ increases. \textcolor{black}{To circumvent this difficulty, oceanographers subsequently pursued} constructions of globally-defined \textcolor{black}{density variables} valid for all pressures, \textcolor{black}{generally referred to as approximately neutral surfaces (ANS)}, the most widely used being: 1) \citet{Reid1971} patched potential density (PPD), which uses a discrete set of vertically stacked potential densities $\sigma_k$ referenced to a discrete set of reference pressures $p_k, k=1,\cdot,N$ spanning the full range of pressures, `patched' at the transition pressures $(p_k+p_{k+1})/2$ for instance; 2) \citet{Jackett1997} empirical neutral density variable $\gamma^n$, \textcolor{black}{which they proposed as a continuous analogue of PPD}, whose iso-surfaces are made up of parcels \textcolor{black}{satisfying the neutrality property (\ref{neutral_path}) as much as feasible}. 

\par 
Nowadays, empirical ANS are more generally envisioned as \textcolor{black}{mathematically well defined} surfaces everywhere as perpendicular as feasible to the \textcolor{black}{non integrable} dianeutral vector ${\bf n} = \alpha \nabla \theta - \beta \nabla S$, this non integrability being generally attached to the non vanishing of the neutral helicity $H_N={\bf n}\cdot (\nabla \times {\bf n}) \ne 0$ \citep{McDougall1988b,Stanley2019a}, where $\alpha = -\rho_{\theta}/\rho$ and $\beta = \rho_S/\rho$ are the thermal expansion and haline contraction coefficients respectively. \textcolor{black}{As a result, the norm $|{\bf n} \times {\bf n}_r^{ans}|$ of the cross product between ${\bf n}$ and any vector ${\bf n}_r^{ans} $ normal to an ANS cannot vanish in general. ANS constructions have been based on treating $|{\bf n} \times {\bf n}_r^{ans}|$ or related quantities as an error (which seems to be a mathematical abuse of the term, since ${\bf n}$ cannot define the `true' value of ${\bf n}_r^{ans}$, even in principle) to be minimised as part of some global ad-hoc and heuristic global optimisation strategy, which has led to widely different and} incompatible approaches; thus, \citet{Eden1999} advocate the use of a purely material density variable $\gamma(S,\theta)$, \citet{deSzoeke2000b} and \citet{Stanley2019a} advocate the use of orthobaric density $\gamma(\rho,p)$, a function of density and pressure only, while Prof. McDougall and his group advocate the use of a hybrid density variable $\gamma(x,y,S,\theta,p)$ or $\gamma(x,y,S,\theta)$ \citep{Jackett1997,Klocker2009,Lang2020,Stanley2021}. 

\par 

\textcolor{black}{Surprisingly, the theoretical justification for the neutral directions appears to remain primarily rooted in \citet{Sverdrup1942} original heuristic two-parcel argument, which in its modern interpretation is viewed as a discrete description of the directions perpendicular to ${\bf n}$. Thus, apart from scarce iconoclastic but inconclusive studies such as \citet{Nycander2011} and \citet{Tailleux2016a,Tailleux2016b}, the neutral directions have neither been really challenged nor given more rigorous foundations in their 80 years of existence; they also appear to have generated little scientific debate or new ideas, except perhaps for \citet{McDougall1987} re-interpretation of the neutral directions as the directions along which the interchange of two parcels does not experience restoring buoyancy forces or a brief but intense altercation between \citet{McDougall2017} and \citet{Tailleux2017}. Likewise, there is only scarce and inconclusive experimental or observational evidence in support or against ANS, such as \citet{Pingree1972}, who found the spread of $\theta/S$ properties to be reduced over neutral surfaces as compared to over selected potential density surfaces, or \citet{vanSebille2011}, who found $\sigma_2$ to outperform $\sigma_0$ and $\gamma^n$ for the tracing of the Labrador seawater from its formation regions to the Abaco line in the Gulf Stream area. In the context of ocean modelling, the neutral directions have been commonly accepted as the directions to be used in \citet{Redi1982} rotated diffusion tensor \citep{Griffies1998b,Shao2020} to reduce the Veronis effect \citep{Veronis1975} that has plagued earlier models, e.g., \citet{Boning1995}. Whether doing so actually succeeds is unknown, however, because how to test and evaluate whether neutral rotated diffusion tensors cause spurious diapycnal mixing or not has remained unclear. }

\textcolor{black}{
In this paper, we show how to embed the theory of lateral stirring and lateral stirring surfaces into the APE-based dynamical study of the full Navier-Stokes equations for compressible seawater, which should enable modellers and theoreticians to develop mathematical and numerical models for its study that can lead to testable predictions, while also making the topic more enticing so as to attract a new generation of scientists who can further help expanding on the currently limited scientific debate. Physically, our approach assumes that lateral stirring represents the notional form of stirring that minimally perturb the APE of the oceans, similarly as \citet{Sverdrup1942}. Indeed, this seems to be the only way to connect the theory of lateral stirring to the current theory of turbulent stratified mixing, which defines diapycnal mixing in terms of the APE dissipation rate $\varepsilon_p$ and regards the Lorenz reference density (LRD) surfaces as the dynamically relevant surfaces for measuring the changes in APE in a fluid characterising vertical stirring, e.g., \citet{Winters1995}. In the past few years, the LRD surfaces have been demonstrated empirically to accurately mimic the $\gamma^n$ surfaces in most of the oceans \citep{Tailleux2016b,Tailleux2021}, which provides strong support for the idea that such surfaces are linked to the lateral stirring surfaces in some way, which this paper will aim to clarify. Until relatively recently, how to use APE theory rigorously for a compressible ocean with a realistic nonlinear equation had seemed out of reach, but following rapid progress over the past decade, \citep{Tailleux2013b,Saenz2015,Tailleux2018}, APE theory has become available as a local concept for general compressible multi-component stratified fluids useful to tackle concrete scientific questions, e.g., \citet{Novak2017}, \citet{Harris2022}.}

\par 
\textcolor{black}{In section \ref{lateral_stirring_surfaces}, we first reformulate \citet{Sverdrup1942} heuristic two-parcel argument in terms of energetics and review known results about oceanic APE and the physics of thermobaricity that are key for correctly interpreting the result. This serves to establish that isoneutral lateral stirring is fundamentally coupled to vertical stirring and that lateral stirring in the oceans inevitably give rise to a new type of forces, called thermobaric forces, regardless of the lateral stirring directions considered. These results establish that the LRD surfaces and ANS should be regarded as distinct surfaces describing two different forms of lateral stirring, the comparison between the two types of surfaces shedding light on the regions where thermobaric forces are too large to ignore. In section \ref{neutral_directions}, we show how to use APE theory to understand how to derive the relevant neutral directions directly from the Navier-Stokes equations for compressible seawater. These directions are found to be the directions perpendicular to an APE-based form of the P vector previously identified by \citet{Nycander2011}. Likewise, the differences between the P-neutrality thus defined and standard N-neutrality can also serve to identify where thermobaric forces are too large to ignore. Section \ref{conclusions} summarises our results and discusses possible future directions. 
}

\section{Thermobaric coupling of lateral and vertical stirring in seawater} 

\label{lateral_stirring_surfaces} 

\subsection{Two-parcel energetics characterisation of stirring} 

\textcolor{black}{To examine the consequences of defining lateral stirring as the notional form of stirring that minimally perturb the APE of the oceans, which is exact in a simple fluid, let us first explicitly estimate the potential energy cost of the adiabatic and isohaline permutation of two fluid parcels, which \citet{Sverdrup1942} did not explicitly discuss. Because thermobaricity causes colder parcels to be more compressible than warmer parcels \citep{Fofonoff1998}, internal energy and compressible effects must play a key role that needs to be elucidated and discussed. This motivates us to use specific enthalpy $h(S,\theta,p)$ as a proxy for potential energy, e.g., \citet{Eden2015,Tailleux2015}}. The predicted potential energy cost of the two parcels exchange is thus
\begin{equation}
\begin{split}
     \Delta E  = & h(S_1,\theta_1,p_2) - h(S_1,\theta_1,p_1) \\ + & h(S_2,\theta_2,p_1) - h(S_2,\theta_2,p_2) \\
     \approx &  -\Delta \upsilon^{LR} \Delta p
      \approx -\frac{1}{\overline{\rho}}[ \overline{\alpha} \Delta \theta - 
      \overline{\beta} \Delta S ] \Delta p ,
\end{split} 
      \label{two_parcel_energy_cost} 
\end{equation}
e.g. \citet{Tailleux2016a}, where $\upsilon^{LR} = \upsilon(S,\theta,\overline{p})$ denotes the so-called `locally-referenced specific volume', $\overline{p} = (p_1+p_2)/2$, and $\overline{\alpha}$ and $\overline{\beta}$ are the thermal expansion and haline contraction coefficients defined in terms of the mean values $\overline{S} = (S_1+S_2)/2$, $\overline{\theta}=(\theta_1+\theta_2)/2$, and $\overline{p}$, \textcolor{black}{while $\Delta (\cdot) = (\cdot)_2 - (\cdot)_1$}. 

\par 

\textcolor{black}{For a simple fluid ($\Delta S = 0$), (\ref{two_parcel_energy_cost}) clearly shows that the adiabatic permutations taking place along a single isentropic surfaces $\theta = \theta_1 = \theta_2$ satisfy $\Delta E = 0$ and can indeed be characterised as leaving the (available) potential energy unaffected. Conversely, the adiabatic permutations involving parcels belonging to two different isentropic surfaces $(\Delta \theta \ne 0)$ must in general entail the deformation of such surfaces with attendant changes in APE $(\Delta E \ne 0)$ (excluding the degenerate isobaric case $\Delta p = 0$), which is the signature of vertical stirring. } 
\textcolor{black}{In that case, the role of buoyancy forces can be explicitly revealed by rewriting} (\ref{two_parcel_energy_cost}) in the form 
\begin{equation}
     \Delta E \equiv \Delta E^{vertical} \approx g \overline{\alpha} \frac{\Delta \theta}{\Delta z} \Delta z^2 \approx \overline{N}^2 \Delta z^2 , 
     \label{vertical_energy_cost_ter}
\end{equation}
where $\Delta z$ is a vertical displacement such that $\Delta p \approx -\overline{\rho} g \Delta z$ and $g \overline{\alpha} \Delta \theta/\Delta z \approx \overline{N}^2$, which predicts buoyancy forces to scale as $\overline{N}^2 \Delta z$, as expected.

\subsection{Lateral stirring and thermobaric forces in seawater}

\textcolor{black}{In the general case, Eq. (\ref{two_parcel_energy_cost}) shows that adiabatic and isohaline permutation of two fluid parcels that minimally perturb the potential energy (i.e., satisfying $\Delta E = 0$) are those satisfying $\overline{\alpha} \Delta \theta - \overline{\beta} \Delta S = 0$ (barring again the degenerate isobaric case), which corresponds to lateral stirring along the standard neutral directions locally perpendicular to ${\bf n} = \alpha \nabla \theta - \beta \nabla S$. While this agrees with standard thinking, it is important to realise that this does not in itself guarantee that isoneutral stirring is necessarily physically realisable, which does not appear to have been previously pointed out. Indeed, isoneutral stirring in seawater differs significantly from that in a simple fluid, in that Eq. (\ref{two_parcel_energy_cost}) shows that it must involve compensating energy changes 
\begin{equation}
      \frac{\overline{\alpha} \Delta \theta \Delta p}{\overline{\rho}} 
      = \frac{\overline{\beta} \Delta S \Delta p}{\overline{\rho}} ,
      \label{compensating_changes} 
\end{equation} 
associated with the deformation of the density-compensated temperature/salinity fields, as $\Delta \theta \ne 0$ and $\Delta S \ne 0$ in general. As we show below, such deformations will in general give rise to both buoyancy and thermobaric forces, so that for isoneutral stirring to achieve a net zero energy cost, work against one force needs to be compensated exactly by work against the other force; isoneutral stirring is physically realisable only if this compensation can actually occur in Nature. 
}
\par 
\textcolor{black}{ 
To shed light on the issue, it is useful to examine the various ways in which the compensating energy changes characterising zero energy cost stirring can be understood. First, let us show that $\Delta E=0$ implies compensating changes between gravitational potential energy (GPE) and internal energy (IE). Indeed, once the two parcels have switched position, the colder parcel will occupy a smaller volume than the warmer parcel that it replaces and vice versa (recall that thermobaricity causes colder parcels to be more compressible than warmer parcels). As a result, the water column above the colder parcel will slightly contract while that over the warmer parcel will slightly expand, thus implying net changes in both GPE and IE. As shown by \citet{Reid1981}, it is this property of thermobaricity that makes available internal energy (AIE) negative in seawater and a significant fraction (up to $40\%$) of the total APE \citep{Huang2005,Tailleux2015}. To confirm this mathematically, let us establish that the change in internal energy $\Delta U = u(S_1,\theta_1,p_2) - u(S_1,\theta_1,p_1) + u(S_2,\theta_2,p_1)-u(S_2,\theta_2,p_2)$ is non-vanishing in the case $\Delta E = 0$. Proceeding similarly as for (\ref{two_parcel_energy_cost}) and using the fact that $\partial u/\partial p |_{S,\theta} = -p \partial \upsilon/\partial p|_{S,\theta}$ yields
\begin{equation}
\begin{split} 
       \Delta U & = \int_{p_1}^{p_2} \left [ \frac{\partial u}{\partial p}
       (S_1,\theta_1,p') - \frac{\partial u}{\partial p}(S_2,\theta_2,p')
       \right ]\,{\rm d}p'  \\ 
         & = -\int_{p_1}^{p_2} p' \left [ \frac{\partial \upsilon}{\partial p}
         (S_1,\theta_1,p') - \frac{\partial \upsilon}{\partial p}(S_2,\theta_2,p')
         \right ] \,{\rm d}p'  \\
         & \approx \overline{p} [ \overline{\upsilon}_{pS} \Delta S 
          + \overline{\upsilon}_{p\theta} \Delta \theta ]  \Delta p .
          \label{IE_cost} 
\end{split} 
\end{equation}
Now, the condition (\ref{compensating_changes}) characterising isoneutral stirring may also be written
in the form $\overline{\upsilon}_S \Delta_n S + \overline{\upsilon}_{\theta} \Delta_n \theta = 0$ (the suffix `n' indicating that the $\Delta$ quantities are estimated along a locally-referenced potential density surface), which if used to eliminate $\Delta_n S$ in (\ref{IE_cost}) leads to 
\begin{equation}
       \Delta U = \frac{\overline{p}}{\overline{\upsilon}_S} 
       ( \overline{\upsilon}_S \overline{\upsilon}_{p\theta} 
       - \overline{\upsilon}_{ps} \overline{\upsilon}_{\theta} ) \Delta_n 
       \theta \Delta_n p = \overline{T}_b
       \Delta_n \theta \Delta_n p \times \frac{\overline{p}}{\overline{\rho}}  
       \label{du_thermobaric}
\end{equation}
where 
\begin{equation}
      T_b = \frac{\rho}{\upsilon_S} \left ( 
      \frac{\partial \upsilon}{\partial S} 
      \frac{\partial^2 \upsilon}{\partial \theta \partial p} 
      - \frac{\partial \upsilon}{\partial \theta} 
      \frac{\partial^2 \upsilon}{\partial S \partial p} \right ) \\ 
      = \frac{\partial \alpha}{\partial p} 
      - \frac{\alpha}{\beta} \frac{\partial \beta}{\partial p} 
    = \beta \frac{\partial}{\partial p} \left ( \frac{\alpha}{\beta} \right ),
        \label{thermobaric_parameter} 
\end{equation}
is the so-called thermobaric parameter \citep{McDougall1987c,Tailleux2016a}. Eq. (\ref{du_thermobaric}) confirms that thermobaricity causes $\Delta U$ (and by implication $\Delta GPE$) to be non-vanishing as long as $\Delta_n \theta \ne 0$ and $\Delta_n p \ne 0$.
}
\par 
\textcolor{black}{Next, we seek a dynamical decomposition of $\Delta E$ in terms of the work against buoyancy and thermobaric forces demonstrating the coupling between vertical and lateral stirring. To achieve this, we use a density/spiciness change of variables $(S,\theta) \rightarrow (\gamma(S,\theta),\xi(S,\theta))$ as in \citet{Tailleux2021}, with $\gamma$ describing the LRD surfaces so that $\Delta \gamma \ne 0$ and $\Delta \xi \ne 0$ can be meaningfully interpreted as indicators of vertical and lateral stirring respectively. Thus, rewriting the equation of state for density as $\rho = \rho(S,\theta,p) = \hat{\rho}(\gamma,\xi,p)$ allows us to rewrite the energy cost (\ref{two_parcel_energy_cost}) in the form
\begin{equation}
        \Delta E \approx  \frac{1}{\hat{\rho}^2} 
        \left ( \frac{\partial \hat{\rho}}{\partial \gamma} \Delta \gamma
        + \frac{\partial \hat{\rho}} {\partial \xi} \Delta \xi \right ) \Delta p ,
        \label{gamma_xi_energy_cost} 
\end{equation}
where
\begin{equation}
      \frac{\partial \hat{\rho}}{\partial \gamma} 
      = \frac{1}{J} \frac{\partial (\xi,\rho)}{\partial (S,\theta)} , 
      \qquad
       \frac{\partial \hat{\rho}}{\partial \xi} 
       = \frac{1}{J} \frac{\partial (\rho,\gamma)}{\partial (S,\theta)} 
       \label{gamma_xi_derivatives} ,
\end{equation}
\citep{Tailleux2021}, 
with $J = \partial (\xi,\gamma)/\partial (S,\theta) = \xi_S \gamma_{\theta} - \xi_{\theta} \gamma_S$ the Jacobian of the transformation. Note that to declutter notation we dropped the overbar, but all quantities remain estimated at the parcels' mean values $(\overline{\gamma},\overline{\xi},\overline{p})$. Physically, thermobaricity affects the energy cost via the spiciness derivative $\partial \hat{\rho}/\partial \xi$, which Eq. (\ref{gamma_xi_derivatives}) shows is controlled by the degree of non-neutrality of $\gamma$. In the case $\Delta E =0$, Eq. (\ref{gamma_xi_energy_cost}) implies that
\begin{equation}
       \underbrace{\frac{1}{\hat{\rho}^2} \frac{\partial \hat{\rho}}{\partial \gamma} 
       \Delta_n \gamma \Delta_n p}_{\Delta E_n^{vertical}} 
       \approx - 
       \underbrace{\frac{1}{\hat{\rho}^2} \frac{\partial \hat{\rho}}{\partial \xi} \Delta_n \xi \Delta_n p}_{\Delta E_n^{lateral}}  \ne 0 ,
       \label{lateral_vertical_coupling}
\end{equation}
and establishes that isoneutral stirring requires compensating work between buoyancy forces $\Delta E_n^{vertical}$ and thermobaric forces $\Delta E_n^{lateral}$ that couples lateral and vertical stirring. For more general permutations $(\Delta E \ne 0)$ taking place on an arbitrary quasi-material surface $\sigma(S,\theta) = {\rm constant}$, (\ref{gamma_xi_energy_cost}) may be rewritten as
\begin{equation}
       \Delta E_{\sigma} = 
       \underbrace{\frac{1}{\hat{\rho}^2} \frac{\partial \hat{\rho}}{\partial \gamma} 
       \Delta_{\sigma} \gamma 
       \Delta_{\sigma} p}_{\Delta E^{vertical}_{\sigma}}  + 
       \underbrace{\frac{1}{\hat{\rho}^2} \frac{\partial \hat{\rho}}{\partial \xi} \Delta_{\sigma} \xi 
       \Delta_{\sigma} p}_{\Delta E^{lateral}_{\sigma}} \ne 0 ,
       \label{energy_cost_arbitrary_permutation} 
\end{equation}
with the suffix `$\sigma$' denoting values taken along the iso-$\sigma$ surface. Eq. (\ref{energy_cost_arbitrary_permutation}) shows that lateral stirring on any arbitrary quasi-material surface will in general involve work against both thermobaric and buoyancy forces, except for lateral stirring along the LRD surfaces $(\sigma = \gamma)$ that only involves work against thermobaric forces owing to its decoupling with vertical stirring. Proceeding as for (\ref{vertical_energy_cost_ter}), the expression for $\Delta E_{\sigma}^{lateral}$ suggest that thermobaric forces acting on the iso-surface $\sigma = {\rm constant}$ scale as 
\begin{equation}
         F^{thermobaric}_{\sigma}  \propto 
         \frac{1}{\hat{\rho}^2} \frac{\partial \hat{\rho}}{\partial \xi} 
         |\nabla_{\sigma} \xi | | \nabla_{\sigma} p | \Delta \ell 
         \label{thermobaric_force} 
\end{equation}
$\Delta \ell$ being a lateral displacement, with $\nabla_{\sigma} \xi$ and $\nabla_{\sigma} p$ the iso-$\sigma$ gradients of $\xi$ and $p$ respectively. 
 }
 \par 
\textcolor{black}{Physically, the condition (\ref{lateral_vertical_coupling}) is a key new result of this paper, for it suggests that isoneutral stirring might be impossible to achieve in Nature, contrary to what has been assumed so far. Indeed, (\ref{lateral_vertical_coupling}) states that for isoneutral stirring to be observable, a necessary condition is that one of the buoyancy or thermobaric forces be destabilising, the other stabilising. From the viewpoint of energetics alone, this is not necessarily impossible, at least in principle, as the case $\Delta E_n^{lateral}<0$, $\Delta E_n^{vertical}>0$ could occur as the result of thermobaric instability \citep{Stewart2016,Tailleux2016a}, here associated with the condition $\hat{\rho}_{\xi} \Delta_n \xi \Delta p <0$, while the case $\Delta E_n^{lateral}>0$, $\Delta E_n^{vertical}<0$ could occur as the result of an instability involving buoyancy forces, such as baroclinic or Kelvin-Helmholtz instability. However, there is no guarantee that the energy released by one of the instabilities should necessarily go towards achieving the desired compensation, as Nature may dictate that it should be diverted to a different energy compartment, in which case lateral stirring would end up occurring along non-neutral directions as should also be the case if both of the forces are simultaneously stabilising or destabilising. This suggests that the binary character of seawater makes it possible for lateral stirring in the oceans to explore a wider range of lateral directions than in a simple fluid, as first suggested by \citet{Tailleux2016a}, which casts doubt on the universal physical significance of the standard neutral directions for lateral stirring. Interestingly, the case $\Delta E_n^{lateral}<0, \Delta E_n^{vertical}>0$ describes the hypothetical case whereby the energy released by thermobaric instability would ultimately cause some diapycnal mixing and dispersion at zero energy cost, which appears to be compatible with Prof. McDougall's longstanding view that dianeutral upwelling without a signature in microstructure measurements should exist as the result of the helical character of finite amplitude neutral trajectories \citep{McDougall2003b}. }
\par 

\subsection{Thermobaric forces attached to LRD surfaces} 

\textcolor{black}{As explained above, thermobaric forces are the key ingredient that makes lateral stirring in seawater fundamentally different from that in a simple fluid and are the ultimate cause for the non-existence of neutral surfaces. The existence of such forces, whose necessity follows from the physical considerations developed above, has the important implication of definitely invalidating the ambiguous concept of `locally-referenced potential density (LRPD) surfaces' that has dominated the literature until now. Physically, this is because the role and existence of thermobaric forces can only be revealed when using mathematically well defined physical variables, which the concept of LRPD fails to achieve.}
\par
\textcolor{black}{Here, we examine the properties and parameters controlling the magnitude of the thermobaric forces that `live' on the LRD surfaces. As explained previously, the LRD surfaces play a central role in this paper due to being the lateral stirring surfaces whose deformations measure APE changes and vertical stirring.}  \citet{Saenz2015} define the LRD as 
\begin{equation}
      \rho^{LZ}(S,\theta) = \rho(S,\theta,p_r)
      \label{LRD_definition} 
\end{equation}
where $p_r = p_0(z_r)$, with $z_r$ the reference depth of a fluid parcel defined as a root of the level of neutral buoyancy (LNB) equation
\begin{equation}
       \rho(S,\theta,p_0(z_r)) = \rho_0(z_r) ,
       \label{LNB_equation} 
\end{equation}
\citep{Tailleux2013b}, with $p_0(z)$ and $\rho_0(z) =-g^{-1} dp_0/dz$ the reference pressure and density profiles defining Lorenz reference state of minimum potential energy. Physically, (\ref{LRD_definition}) defines the LRD as a generalised form of potential density referenced to the spatially variable reference pressure $p_r(S,\theta)$. Importantly, (\ref{LNB_equation}) defines $z_r$ as the intersection point of two one-dimensional curves and therefore as a local quantity \textcolor{black}{in the conventional mathematical sense of the term. Eq. (\ref{LNB_equation}) also shows that $z_r$ is parameterically dependent on the globally defined Lorenz reference state, so that any ambiguity in the determination of $\rho_0(z)$ and $p_0(z)$, which in practice may arise from our imperfect knowledge of the ocean stratification or from neglecting its time dependence, will introduce some uncertainty in the value of $z_r$. However, such an issue only matters in concrete applications; in theoretical work, as is the case here, $z_r$ may be assumed to be known exactly at all times without loss of generality. Note also that being parameterically dependent on globally defined quantities does not make $z_r$ a global quantity, contrary to what is sometimes believed. }

\par 

The LNB equation (\ref{LNB_equation}) plays a key role in the local theory of APE, for it encodes all the information about $z_r$. For instance, differentiating it yields
\begin{equation}
     \nabla z_r = \left ( \frac{d\rho_0}{dz}(z_r) + 
     \frac{\rho_0(z_r)g}{c_{sr}^2} \right )^{-1} 
     (\rho_{Sr} \nabla S + \rho_{\theta r} \nabla \theta ) ,
     \label{gradient_zr} 
\end{equation}
where $c_s = \rho_p^{-1/2}(S,\theta,p)$ is the sound speed, while the suffix `r' denotes quantities estimated at the reference pressure $p_r$, which in turns implies
\begin{equation}
      \nabla \rho^{LZ} = -\underbrace{\rho_r \frac{d\rho_0}{dz}(z_r) 
      \left ( \frac{d\rho_0}{dz}(z_r) + 
     \frac{\rho_r g}{c_{sr}^2} \right )^{-1}}_{b_0} {\bf n}_r,
     \label{gradient_rholz} 
\end{equation}
where ${\bf n}_r = \alpha_r \nabla \theta - \beta_r \nabla S$ defines a reference neutral vector, while $\rho_r = \rho_0(z_r)$. Mathematically, (\ref{gradient_zr}) and (\ref{gradient_rholz}) establish that the gradients of $z_r$, $p_0(z_r)$ and $\rho^{LZ}$ are all proportional to ${\bf n}_r$, and therefore that the iso-surfaces of LRD, $z_r$, and $p_r=p_0(z_r)$ all coincide. 

\par 

If we now set \textcolor{black}{$\gamma = \rho^{LZ} (S,\theta) = \rho(S,\theta,p_r)$, using $\xi = \theta$ for simplicity so that $J= -\gamma_S$, the $\xi$ derivative of $\hat{\rho}$ becomes 
\begin{equation}
\begin{split} 
     \frac{\partial \hat{\rho}}{\partial \xi} 
      = &  \rho_{\theta} - \frac{\rho_S}{\gamma_S} \gamma_{\theta} 
      = \rho_S \int_{p}^{p_r} \frac{\partial}{\partial p}
      \left (\frac{\alpha}{\beta} \right) (p')\,{\rm }p' \\
       = & \rho_S \int_{p}^{p_r} \frac{T_b}{\beta} {\rm d}p' 
       \approx \rho T_b  (p_r - p) , 
\end{split} 
\end{equation}
which yields
\begin{equation}
      \Delta E_{LRD}^{lateral} \approx 
      \frac{\overline{T}_b}{\overline{\rho}} \Delta_i \theta 
      \Delta_i p (p_r - \overline{p})  ,
      \label{energy_cost_lateral_stirring} 
\end{equation}
where the suffix `i' is used specifically to refer to isopycnal variations measured on the LRD surfaces.} 
Eq. (\ref{energy_cost_lateral_stirring}) thus predicts that the \textcolor{black}{thermobaric forces acting along the} LRD surfaces are controlled by: a) the thermobaric parameter $T_b$; b) the isopycnal gradient of potential temperature $\Delta_i\theta$; c) the distance from the equilibrium state of rest $p_r - \overline{p}$; \textcolor{black}{ it also shows that like buoyancy forces, thermobaric forces depend on the distance to Lorenz reference state $|p-p_r|$ and hence on the global ocean stratification. Lateral stirring in seawater is therefore slightly non-local as a result. }   

\subsection{Lorenz reference density surfaces versus approximately neutral surfaces}

\textcolor{black}{
Where thermobaric forces are weak enough to be neglected, the above results establish that seawater approximately behaves like a simple fluid and that the LRD surfaces are then sufficiently accurately neutral to be regarded as the appropriate definition of lateral stirring surfaces. Where thermobaric forces are large, however, lateral stirring along the LRD surfaces entails a non-zero energy cost, which means that it can not occur without also interacting with other energy compartments of the system, which might be enough to cause lateral stirring to effectively occur along different directions. It appears therefore necessary to introduce a new type of lateral stirring surface, baptised here Lateral Mixing Surfaces (LMS) as a separate and distinct concept from the LRD surfaces. Physically, the LMS are envisioned as defining physically realisable mixing paths, as per \citet{Foster1976} terminology, without however seeking to imply that these paths should be necessarily defined as they did. Whether the concept of LMS can be meaningfully defined and constructed remains tentative at this stage; for the time being, it is convenient to assume that it coincides with the isopycnal surfaces that oceanographers have been after all along and hence that the various empirical ANS proposed so far in oceanography represent our current best guesses of LMS, although we anticipate based on the results of the following section that this might evolve rapidly in the near future. 
}
\par 
\textcolor{black}{ 
Currently, an empirical ANS can be defined quite generally as a mathematically well defined surface whose degree of non-neutrality $|{\bf n} \times {\bf n}_r^{ans}|$ is determined by means of some heuristic global optimisation problem, where ${\bf n}_r^{ans}$ defines a vector normal to the ANS considered. As a result, ${\bf n}_r^{ans}$ and the lateral stirring directions that it defines depends on the global ocean stratification and are therefore slightly non local in a way that varies from one method to the other. Thus, in \citet{Eden1999}'s approach, this non-local dependence arises from the global elliptic problem used; in \citet{Jackett1997}, it arises from the value of $\gamma^n$ at any point in the oceans being determined by the value of $\gamma^n$ on a reference cast in the Pacific Ocean to which it is neutrally connected; in \citet{deSzoeke2000b,deSzoeke2009} or \citet{Stanley2019a}, it arises from the globally or regionally determined $\theta/S$ relationships entering the construction of orthobaric density and topobaric surfaces. Presumably because oceanographers have generally assumed (erroneously as it turns out) lateral stirring to be local in the oceans, similarly as for a simple fluid, the non-local character of the lateral stirring directions attached to empirical ANS has not received much attention, with $|{\bf n}\times {\bf n}_r^{ans}|$ being commonly regarded as an error and its non-local character as spurious. In our view, however, lateral stirring in seawater must be regarded as slightly non-local due to the existence of thermobaric forces, while $|{\bf n} \times {\bf n}_r^{ans}|$ should be regarded, at least partly, as a physical measure of the magnitude of thermobaric processes rather than just an error.
} 
\par
\textcolor{black}{How is $|{\bf n} \times {\bf n}_r^{ans}|$ controlled by thermobaricity, spiciness, or any other parameter(s) determining it remains poorly understood and is generally not addressed in the ANS literature. In contrast, $|{\bf n}\times {\bf n}_r|$ can be easily and explicitly evaluated in closed mathematical form for the LRD surfaces. To show this, let us take the cross product of ${\bf n} = \alpha \nabla \theta - \beta \nabla S$ with ${\bf n}_r = \alpha (S,\theta,p_r) \nabla \theta - \beta(S,\theta,p_r)\nabla S = \alpha_r \nabla \theta - \beta_r \nabla S$, defined as a reference neutral vector perpendicular to LRD surfaces. After some straightforward algebra, making use of (\ref{thermobaric_parameter})'s definition of $T_b$, one may show that:
\begin{equation}
\begin{split} 
     {\bf n} \times {\bf n}_r &  = \beta \beta_r 
     \left (\frac{\alpha}{\beta} 
     - \frac{\alpha_r}{\beta_r} \right )  
     \nabla S \times \nabla \theta \\
      & = \beta \beta_r \int_{p_r}^{p} \frac{T_b}{\beta} (S,\theta,p')\,{\rm d}p' \, \nabla S \times \nabla \theta \\
     & =  \frac{\beta \beta_r \overline{T}_b(p-p_r)}{\overline{\beta}}
     \nabla S \times \nabla \theta 
     \label{nnr_parallelism} 
\end{split} 
\end{equation}
for suitably defined mean values $\overline{T}_b$ and $\overline{\beta}$. 
Like (\ref{energy_cost_lateral_stirring}) or neutral helicity, Eq. (\ref{nnr_parallelism}) reveals the three physical limits for which exact neutrality can be achieved, namely: 1) vanishing thermobaricity $T_b \rightarrow 0$; 2) state of rest, $|p-p_r| \rightarrow 0$; 3) coincidence of isothermal and isohaline surfaces $|\nabla S \times \nabla \theta| \rightarrow 0$. Since $\nabla S \times \nabla \theta$ is a local parameter, the non-local dependence of $|{\bf n}\times {\bf n}_r|$ on the ocean stratification is primarily via $|p-p_r|$ measuring the distance from Lorenz reference state, which is the same parameter controlling the magnitude of vertical buoyancy forces involved in the vertical stirring process.  }

\textcolor{black}{In the ANS literature, it is the magnitude of $H_N$ that has been generally regarded as the main measure of thermobaricity determining the range of possible behaviours of empirical ANS. In particular, \citet{Jackett1997} have argued that oceanic values of $H_N$ are sufficiently small that the inherent ambiguity attached to the density values of any empirical ANS can be expected to remain ``below the present instrumentation error in density''. Such a conclusion is important, because if true, it suggests that thermobaric forces only matter in localised regions of the oceans and hence that empirical ANS should only marginally differ from the LRD surfaces in most of the oceans. As it happens, this is consistent with the results of \citet{Tailleux2016b}, who found the LRD surfaces as described by $\gamma^T$ to accurately coincide with \citet{Jackett1997} $\gamma^n$ surfaces almost everywhere outside the Southern Ocean, where $\gamma^T$ was defined as an empirically pressure-corrected form of LRD 
\begin{equation}
     \gamma^T(S,\theta) = \rho^{LZ}(S,\theta) - f_n(p_r) ,
     \label{gammaT_definition} 
\end{equation}
with $f_n(p_r)$ a polynomial function of $p_r$ empirically fitted to make $\gamma^T$ mimic $\gamma^n$ as much as feasible. 
} 
\textcolor{black}{This is also consistent with the results of \citet{Tailleux2021}, who repeated the same comparison using $\gamma^T_{analytic}$, a modified form of $\gamma^T$ based on an analytical representation of Lorenz reference state. Because $\gamma^n = \gamma^n(S,T,p,x,y)$ is a priori a function of location, the comparisons between $\gamma^T$ or $\gamma^T_{analytic}$ and $\gamma^n$ have been primarily carried out in physical space so far. However, both \citet{McDougall2005b} and \citet{Lang2020} have estimated the non-material effects arising from the $(x,y)$ dependence of $\gamma^n$ to be negligible, suggesting that the latter might be close to be quasi-material. To test this, we constructed a new quasi-material interpolant $\gamma_{material}^n = \gamma_{material}^n(S,\theta)$ of $\gamma^n$ to compare it with $\gamma^T_{analytic}$ directly in $(S,\theta)$ space, obtained by `feeding' the specialised Matlab routine {\tt scatteredInterpolant} with values of $\gamma^n$, $S$ and $\theta$ from \citet{Gouretski2004} climatology. In contrast to  \citet{McDougall2005b}'s poor material approximant $\gamma^a(S,\theta)$, constructed in terms of rational functions, $\gamma^n_{material}$ is found accurately approximate both $\gamma^n$ and $\nabla \gamma^n$ nearly everywhere (not shown for lack of space).} 

\begin{figure*}[t]
\centerline{\includegraphics[width=39pc,angle=0]{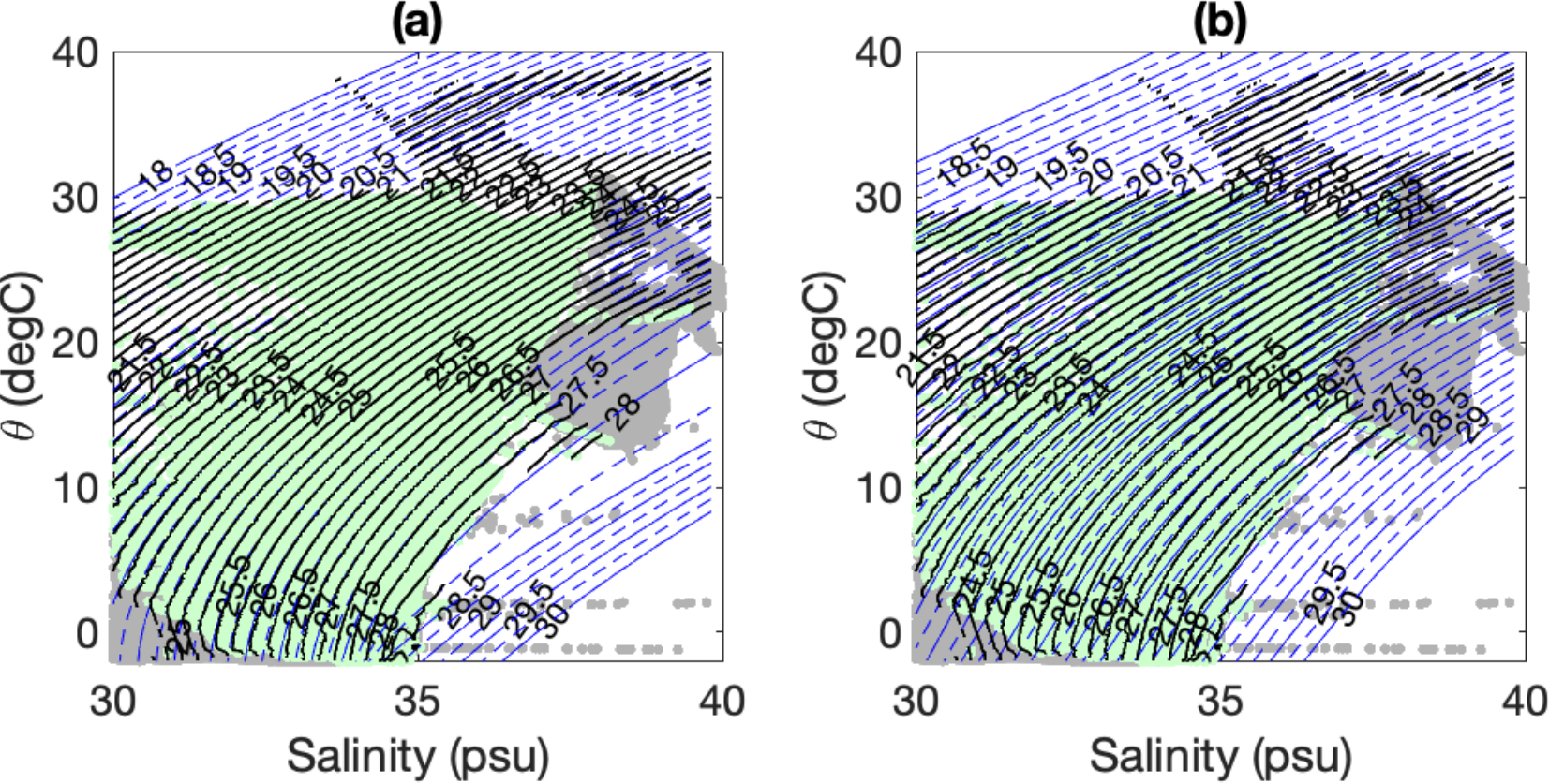}} 
\caption{(a) Comparison between $\gamma^T_{analytic}$ and $\gamma^n_{material}$ (b) and between $\gamma^a$ and $\gamma^n_{material}$. In all panels, the thick black lines indicate the iso-contours of $\gamma^n_{material}$, the green area indicate the subpart of $(S,\theta)$ space where $\gamma^n$ is defined, and the grey area indicate the additional points of the climatology for which $\gamma^n$ is not defined. The black contours line indicate the iso-contours of $\gamma^T_{analytic}$ (a) and $\gamma^a$ (b).} 
\label{f0} 
\end{figure*} 

Our prediction that $\gamma^n_{material}$ and $\gamma^T_{analytic}$ should accurately coincide with each other outside the polar regions is clearly demonstrated in Fig. \ref{f0} (a), and is evidenced by the near perfect coincidence of the black solid lines and black solid/dashed lines almost everywhere except for the coldest and densest waters \textcolor{black}{where $|p-p_r|$ and thermobaric forces are expected to be the largest}. In contrast, Fig. \ref{f0} (b) shows large differences between $\gamma^a$ and $\gamma^n_{material}$, which is consistent with $\gamma^a$ only poorly approximating $\gamma^n$. Note that for plotting purposes, all the values of $\gamma^n_{material}$ insufficiently constrained by data were set to {\tt NaN}, with the green and grey area in both panels representing the parts of the oceans over which $\gamma^n$ is defined and not defined respectively. \textcolor{black}{These results appear to confirm, therefore, that the LRD surfaces are able to capture the leading order behaviour of LMS in most of the oceans due to thermobaric forces being large only for the coldest and densest water masses.  }

\section{Extension to the full Navier-Stokes equations} 
\label{neutral_directions} 

\subsection{APE-theory and optimal form of momentum balance} 

\textcolor{black}{We now show how to extend the two-parcel based energetics considerations developed in the previous section to the full Navier-Stokes equations. To that end, it proves crucial to write the momentum balance equations in their thermodynamic or Crocco-Vazsonyi \citep{Crocco1937,Vazsonyi1945} form, }
\begin{equation}
     \frac{\partial {\bf v}}{\partial t} + 
     \boldsymbol{\omega}_a \times {\bf v} + \nabla {\cal B}_h 
      = {\bf P}_h + {\bf F} ,
      \label{EGTBF_NSE} 
\end{equation}
as it is the form that most naturally displays how thermodynamics and energetics constrain the forces acting on fluid parcels, which is what we are after. 
Eq. (\ref{EGTBF_NSE}) is
obtained from (\ref{PGVBF_NSE}) by making use of the total differential for specific enthalpy ${\rm d}h = T{\rm d}\eta + \mu {\rm d}S + \rho^{-1} {\rm d}p$ and of the identity $({\bf v}\cdot \nabla ){\bf v} = (\nabla \times {\bf v}) \times {\bf v} + \nabla ({\bf v}^2/2)$, where $\boldsymbol{\omega}_a = \nabla \times {\bf v} + 2 \boldsymbol{\Omega}$ is the absolute vorticity, ${\cal B}_h$ and ${\bf P}_h$ being given by 
\begin{equation}
      {\cal B}_h = \frac{{\bf v}^2}{2} + h  + \Phi , \qquad 
      {\bf P}_h = T \nabla \eta + \mu \nabla S ,
      \label{ph_definition} 
\end{equation}
where the quantity $h+\Phi$ is called the static energy in the atmospheric literature. 

\par 
\textcolor{black}{In a simple fluid, the vector ${\bf P}_h = T \nabla \eta + \mu \nabla S$ in (\ref{EGTBF_NSE}) and (\ref{ph_definition}) reduces to ${\bf P} = T \nabla \eta$ and is naturally perpendicular to the lateral stirring surfaces.
It is therefore the force of most obvious interest for the present purposes. Unfortunately, ${\bf P}_h$ in seawater does not appear to be perpendicular to any recognisable form of isopycnal surfaces. Before concluding that the approach does not work in seawater, however, it is crucial to recognise that neither the thermodynamic form of momentum balance (\ref{EGTBF_NSE}) nor the definitions of ${\cal B}_h$ and ${\bf P}_h$ are unique, because any transformation of the form
\begin{equation}
     {\cal B}_h \rightarrow {\cal B}_h - B_0(\eta,S) , \qquad
     {\bf P}_h \rightarrow {\bf P}_h - \left ( \frac{\partial B_0}{\partial 
     \eta} \nabla \eta + \frac{\partial B_0}{\partial S} \nabla S \right ) ,
\end{equation} 
provides mathematically equivalent alternative forms of momentum balance that are also thermodynamic in character, with $B_0(\eta,S)$ any arbitrary quasi-material function of $\eta$ and $S$. The question, therefore, is whether a best choice of $B_0(\eta,S)$ exist that can give us a modified form of ${\bf P}_h$ with the desired properties? Given the central role played by APE theory in the previous section, we assume that the answer is positive and that $B_0$ is related to the background value of the Bernoulli function in Lorenz reference state. To show that this leads to a physically acceptable theory, we thus decompose} ${\cal B}_h = {\cal B}_a + {\cal B}_r$ as the sum of its dynamically active and inert parts respectively, which leads us to introduce the more dynamically relevant P vector ${\bf P}_a = {\bf P}_h - \nabla {\cal B}_r$. As shown below, ${\bf P}_a$ is found to have the desired property of being approximately perpendicular to both the LRD and conventional neutral surfaces, where ${\cal B}_r$ is defined as the value of ${\cal B}_h$ in Lorenz reference state, viz., 
\begin{equation}
       {\cal B}_r = h(\eta,S,p_0(z_r)) + g z_r ,
\end{equation}
whose gradient is 
\begin{equation}
\begin{split} 
     \nabla {\cal B}_r  = & T_r \nabla \eta + \mu_r \nabla S
      + g \left ( 1 - \frac{\rho_0(z_r)}{\rho(\eta,S,p_0(z_r))} \right ) \nabla z_r \\
      = & T_r \nabla \eta + \mu_r \nabla S ,
\end{split} 
\end{equation}
the simplification being due to $z_r$ satisfying the LNB equation (\ref{LNB_equation}). Subtracting $\nabla {\cal B}_r$ from both sides of Eq. (\ref{EGTBF_NSE}) then yields
\begin{equation}
      \frac{\partial {\bf v}}{\partial t} 
      + \boldsymbol{\omega}_a \times {\bf v} 
      + \nabla {\cal B}_a = {\bf P}_a + {\bf F} 
      \label{EGTBF_with_bapa} 
\end{equation}
where ${\cal B}_a$ and ${\bf P}_a$ may be written in the form
\begin{equation}
     {\cal B}_a = \frac{{\bf v}^2}{2} + h + \Phi - {\cal B}_r = 
     \frac{{\bf v}^2}{2} + \Pi + \frac{p-p_0(z)}{\rho}, 
     \label{available_bernoulli} 
\end{equation}
\begin{equation}
     {\bf P}_a  = \frac{\partial \Pi}{\partial \eta} \nabla \eta 
     + \frac{\partial \Pi}{\partial S}     \nabla S = 
     (T-T_r)\nabla \eta + (\mu-\mu_r) \nabla S ,
     \label{pa_expression} 
\end{equation}
where $\Pi = h(\eta,S,p) - h(\eta,S,p_0(z_r)) + g (z-z_r) + (p_0(z)-p)/\rho$ is the potential energy density of a compressible two-component stratified fluid, e.g., \citet{Tailleux2018}, which may be regarded as 
the sum of available compressible energy $\Pi_1$ and APE density $\Pi_2$, 
\begin{equation}
\begin{split} 
     \Pi_1  = h(\eta,S,p) & - h(\eta,S,p_0(z))  + \frac{p_0(z)-p}{\rho} \\
     & \approx \frac{(p-p_0(z))^2}{2\rho_b^2 c_{sb}^2} 
\end{split} 
\end{equation}
\begin{equation}
\begin{split} 
      \Pi_2  = h(\eta,S,p_0(z)) & - h(\eta,S,p_0(z_r))  +  g (z-z_r) \\ 
      & \approx \frac{N_r^2 (z-z_r)^2}{2} ,
\end{split} 
\end{equation}
where $N_r^2$ is the reference value of the squared buoyancy frequency, the suffix `b' denoting values evaluated at the pressure $p_0(z)$, i.e., $\rho_b = \rho(\eta,S,p_0(z))$.

\textcolor{black}{For simplicity, we ignore the time dependence of $\rho_0(z)$ and $p_0(z)$, as it only affects ${\bf P}_a$ in a parameteric way. For details about how to obtain (\ref{pa_expression}) and the precise meaning of thermodynamic derivatives see Appendix B}. Note here that the quantity 
\begin{equation}
\begin{split} 
     M = & h+\Phi - {\cal B}_r = \Pi + \frac{p-p_0(z)}{\rho} \\
     = & h(\eta,S,p)-h(\eta,S,p_0(z_r)) + g(z-z_r) 
     \label{montgomery_potential} 
\end{split} 
\end{equation}
represents a generalisation of the well known Montgomery potential \citep{Montgomery1937} or acceleration potential \citep{Wexler1941}, see \citet{Stanley2019b} for a recent discussion.

\subsection{Link between ${\bf P}_a$, LRD surfaces, and N-neutral vector} 

To establish that ${\bf P}_a$ is approximately parallel to ${\bf n}_r$ and ${\bf n}$ as claimed above, the simplest is to switch variables and to regard specific enthalpy $h = \hat{h}(S,\theta,p)$ as a function of $(S,\theta,p)$ so as to write its total differential in the form
\begin{equation}
       {\rm d}\hat{h} = \frac{\partial \hat{h}}{\partial \theta}
        {\rm d}\theta + \frac{\partial \hat{h}}{\partial S} {\rm d}S 
        + \hat{\upsilon}\, {\rm d}p . 
\end{equation}
The Maxwell relationships (i.e., the equality of the cross-derivatives), viz.,
\begin{equation}
     \frac{\partial^2 \hat{h}}{\partial \theta \partial p} = 
     \frac{\partial \hat{\upsilon}}{\partial \theta} = 
     \frac{\hat{\alpha}}{\hat{\rho}} , 
     \qquad 
     \frac{\partial^2 \hat{h}}{\partial S \partial p} = 
     \frac{\partial \hat{\upsilon}}{\partial S} 
     = - \frac{\hat{\beta}}{\hat{\rho}} ,
\end{equation}
then allow one to rewrite ${\bf P}_a$ as
\[
       {\bf P}_a = \int_{p_r}^p 
        \frac{\hat{\alpha}}{\hat{\rho}} (S,\theta,p')\,{\rm d}p' 
        \nabla \theta - \int_{p_r}^p 
        \frac{\hat{\beta}}{\hat{\rho}}(S,\theta,p')\,{\rm d}p' 
        \nabla S 
\]
\begin{equation} 
    = \frac{(p-p_r)}{\overline{\rho}} (\overline{\alpha} \nabla \theta 
    - \overline{\beta} \nabla S ) = 
    \frac{p-p_r}{\overline{\rho}} \overline{{\bf n}} ,
\end{equation}
where $\overline{\alpha}$ and $\overline{\beta}$ are
\begin{equation}
      \overline{\alpha} 
      = \frac{\overline{\rho}}{p-p_r} \int_{p_r}^p 
        \frac{\hat{\alpha}}{\hat{\rho}} (S,\theta,p')\,{\rm d}p' , 
        \label{mean_alpha}
\end{equation}
\begin{equation}
      \overline{\beta}  =  \frac{\overline{\rho}}{p-p_r} 
      \int_{p_r}^p \frac{\hat{\beta}}{\hat{\rho}} 
      (S,\theta,p')\,{\rm d}p' ,
      \label{mean_beta} 
\end{equation}
while $\overline{\rho}$ is a representative mean value of $\rho$ over $[p,p_r]$. Using a simple trapezoidal scheme to approximate the integrals in (\ref{mean_alpha}) and (\ref{mean_beta}), as well as the Boussinesq approximation, shows that at leading order
\begin{equation}
       \overline{\bf n} \approx  \frac{1}{2} 
        ( {\bf n} + {\bf n}_r )  
        \label{sought_for_result}
\end{equation}
Eq. (\ref{sought_for_result}) is the sought-for result that establishes that ${\bf P}_a$ is in general intermediate between ${\bf n}$ and ${\bf n}_r$. If $|p-p_r|$ is small, the directions defined by ${\bf P}_a$, ${\bf n}$ and ${\bf n}_r$ should all approximately coincide, but start to grow further apart as $|p-p_r|$ increases. \textcolor{black}{\citet{Nycander2011} obtained a similar result for his ${\bf P}$ vector in the particular case $p_r=0$.} 
\par

\textcolor{black}{To improve on our two-parcel based prediction of thermobaric forces (\ref{thermobaric_force}) in the Boussinesq limit $\Pi_1 \rightarrow 0$, $\Pi \approx \Pi_2$ (as clarified in next section), we note from Appendix B that the gradient of $\Pi_2$ may be written 
\begin{equation}
     \nabla \Pi_2 = {\bf P}_{a2} - b {\bf k} = {\bf P}_{a2}^{(i)} 
     + {\bf P}_{a2}^{(d)} 
     - b {\bf k} 
     \label{gradient_pi2_decomposition}
\end{equation}
where $b = -g(1-\rho_0(z)/\rho_b)$ defines the standard buoyancy force relative to Lorenz reference density profile, while ${\bf P}_{a2}^{(i)}$ and ${\bf P}_{a2}^{(d)}$ represent the component of ${\bf P}_{a2}$ perpendicular and parallel to ${\bf n}_r$ respectively. Using the approximation ${\bf P}_{a2} \approx (p-p_r) ({\bf n}+ {\bf n}_r)/(2\overline{\rho})$ derived above and the fact that ${\bf n}_r^{(i)} =0$ and ${\bf n}^{(i)} 
= \alpha \nabla_i \theta - \beta \nabla_i S$ by definition, leads to
}
\begin{equation}
      {\bf P}_a^{(i)} \approx \frac{p-p_r}{2 \overline{\rho}} 
      {\bf n}^{(i)} 
              \approx  \frac{(p-p_r)^2}{2\overline{\rho}} 
              T_{br} \nabla_i \theta , 
              \label{thermobaric_forces_pa} 
\end{equation}
where we also used the fact that $\alpha_r \nabla_i \theta  = \beta_r \nabla_i S$ due to density-compensation, and a Taylor series expansion of $\alpha$ and $\beta$ around $p_r$ as before. \textcolor{black}{Eq. (\ref{thermobaric_forces_pa}) shows that the thermobaric forces acting along the LRD surfaces would vanish in all 3 idealised physical limits identified before. In contrast, the thermobaric forces discussed by \citet{deSzoeke2000a} for instance do not a priori vanish in a resting state, which is unphysical. It follows that the use of the spatially variable reference pressure entering APE theory is crucial to construct a physically meaningful description of thermobaric forces that vanish in all 3 idealised limits identified before. It is also useful to remark that all information about both the buoyancy and thermobaric forces is contained in the partial derivatives of $\Pi_2$ as shown by Eqs. (\ref{gradient_pi2_decomposition}) and (\ref{thermobaric_forces_pa}), thus highlighting the fundamental importance of $\Pi_2$ for elucidating all aspects of the problem.}  

\subsection{Energetics significance of ${\bf P}_a$} 

The P-vector ${\bf P}_a$ is of fundamental importance in the present theory as it can be shown to define neutral directions along which stirring leaves the potential energy $\Pi$ approximately unaffected. The associated form of neutrality is called P-neutrality to distinguish it from \citet{McDougall1987} conventional N-neutrality. In the oceans, $\Pi \approx \Pi_2$ as $\Pi_1$ is generally several orders of magnitude smaller than $\Pi_2$ and can be formally neglected in the incompressible limit $c_s \rightarrow +\infty$. If we do so, while also approximating $\rho$ by a constant reference density $\rho_{\star}$ in (\ref{available_bernoulli}-\ref{pa_expression}), yields the following Boussinesq-like approximation
\begin{equation}
     {\cal B}_a \approx  \frac{{\bf v}^2}{2} + \Pi_2 + \frac{p-p_0(z)}{\rho_{\star}} ,
     \label{ba_approximated} 
\end{equation}
\begin{equation}
      {\bf P}_a \approx {\bf P}_{a2} = \frac{\partial \Pi_2}{\partial 
      \eta} \nabla \eta + \frac{\partial \Pi_2}{\partial S} \nabla S =
      (T_b-T_r)\nabla \eta + (\mu_b - \mu_r) \nabla S  ,
       \label{pa_approximated} 
\end{equation}
see Appendix B for details, where it is also shown that the work against ${\bf P}_{a2} $ may be written in the form
\begin{equation}
     {\bf v}\cdot {\bf P}_{a2}  
      = (T_b-T_r) \dot{\eta} + (\mu_b - \mu_r) \dot{S} - \frac{\partial \Pi_2}{\partial t} .
      \label{work_against_p2} 
\end{equation}
This relation implies therefore than in the absence of irreversible mixing, the directions perpendicular to ${\bf P}_{a2} $ define the directions along which stirring leaves $\Pi_2$ unaffected. This relation is analogous to Eq. (21) of \citet{Nycander2011}, reproduced here in local form
\begin{equation}
      {\bf v} \cdot {\bf P}_{N11} = 
      \frac{\partial h^{\ddag}}{\partial \theta} \dot{\theta}
      + \frac{\partial h^{\ddag}}{\partial S} \dot{S}
      - \frac{\partial h^{\ddag}}{\partial t}  ,
\end{equation}
which shows that in the absence of mixing, the directions normal to Nycander P-vector are those along which stirring leaves dynamic enthalpy $h^{\ddag}$ unaffected.
\textcolor{black}{The fact that ${\bf P}_{a2}$ depends on the global ocean stratification through its dependence on Lorenz reference state supports the key hypothesis formulated in this paper that lateral stirring is no longer purely local in seawater because of its thermobarically-induced partial coupling to vertical stirring. }

\subsection{Quantification of P-neutrality versus N-neutrality} 

To shed light on the differences between N-neutrality and P-neutrality, we used the \citet{Gouretski2004} WOCE climatology to understand what observations can tell us about the actual differences between ${\bf P}_a$, ${\bf n}$, $\nabla \gamma^T$ and $\nabla \gamma^n$ in the oceans. To exploit the capabilities of the TEOS-10 Matlab Gibbs Seawater Library (available at {\tt www.teos-10.org}), the practical salinity and in-situ temperature fields were converted into reference composition salinity $S_R$ and Conservative Temperature $\Theta$ by means of the routines {\tt gsw\_SR\_from\_SP} and {\tt gsw\_CT\_from\_t} respectively. 

\par 
\textcolor{black}{The standard N-neutral directions attached to ${\bf n}$ were defined in terms of}
\begin{equation}
     {\bf N}^{\star} = \nabla_{lr} \rho(\Theta,S_R,p_0(z)) , 
\end{equation}
(so that ${\bf N}^{\star} \approx -\rho_0 {\bf n})$, 
using standard second-order centred finite differences, 
with $\nabla_{lr}$ the `locally-referenced' gradient, that is, the gradient calculated by ignoring the pressure dependence. As to the new P-neutral directions attached to ${\bf P}_{a2}$, they were defined in terms of 
\begin{equation}
    \overline{{\bf N}}^{\star} \approx \frac{1}{z-z_r} \left [
    \nabla_{lr} h(\Theta,S_R,p_0(z)) 
    - \nabla_{lr} h(\Theta,S_R,p_0(z_r)) \right ] ,
\end{equation}
\textcolor{black}{(so that $\overline{\bf N}^{\star} \approx - {\bf P}_{a2}/(g(z-z_r))$)}, 
the specific enthalpy $h=h(S_R,\Theta,p)$ being estimated using {\tt gsw\_enthalpy\_CT\_exact}.

\begin{figure*}[t]
\centerline{\includegraphics[width=33pc,angle=0]{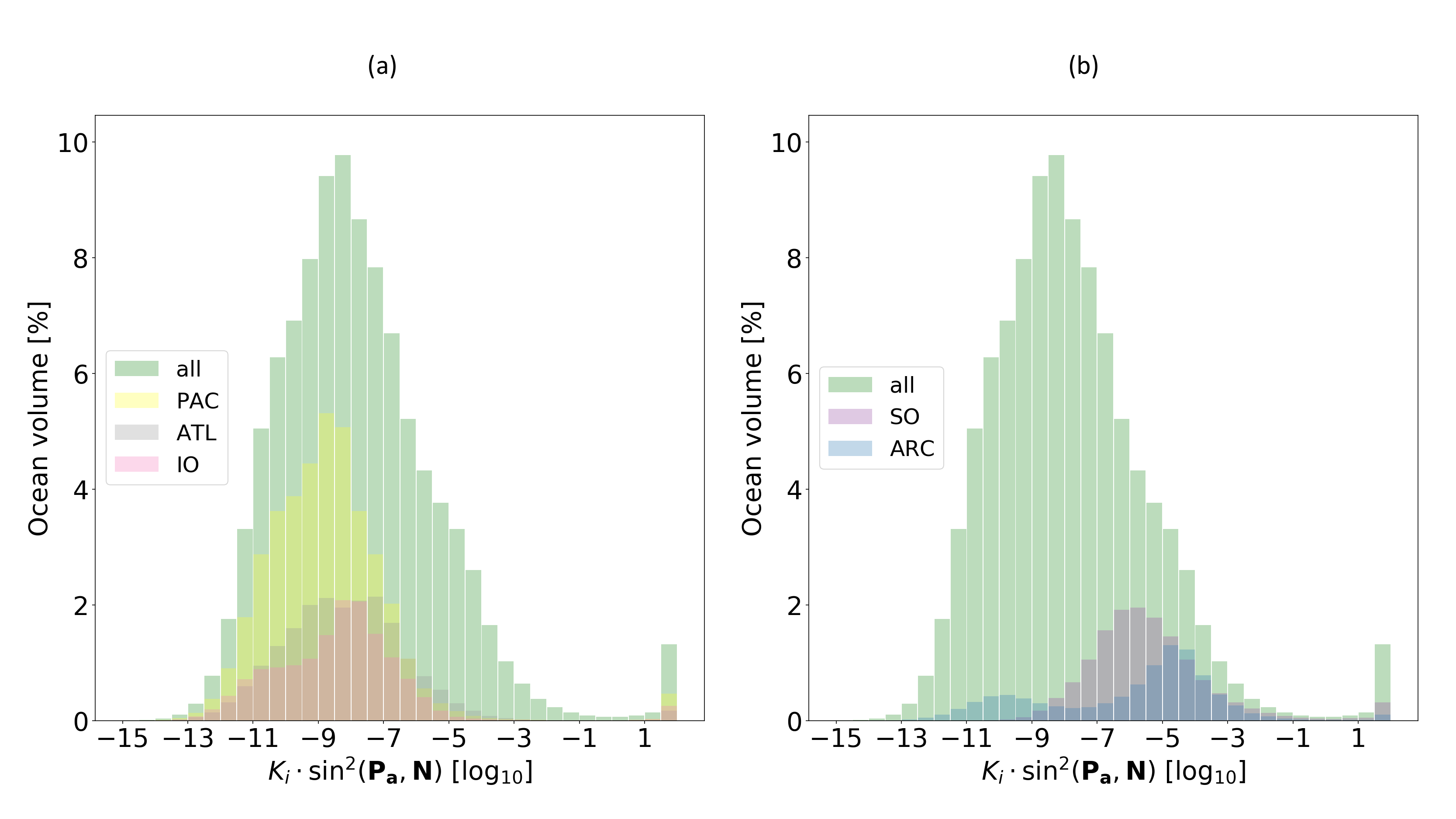}}
\caption{Probability distribution function (pdf) of the decimal logarithm of the effective diapycnal diffusivity-like metric measuring the angle between ${\bf P}_{a2}$ and ${\bf N}$ for the Pacific (PAC), Atlantic (ATL), and Indian (IO) oceans (left panel) versus for the polar oceans (right panel), the statistics for the whole ocean being also shown in the background in both panels. (SO = Southern Ocean, ARC = Arctic Ocean)} 
\label{f1} 
\end{figure*} 

One conventional metric to quantify the differences between two directions ${\bf A}$ and ${\bf B}$ is in terms of the notional effective diffusivity 
\begin{equation} 
    K_f({\bf A},{\bf B}) = K_i \sin^2{\widehat{({\bf A},{\bf B}})} 
    = K_i \frac{|{\bf A} \times {\bf B} |^2}{|{\bf A}|^2
    |{\bf B}|^2} 
    \label{kf_definition} 
\end{equation} 
e.g., \citet{Hochet2019}, with $K_i = 1000\,{\rm m^2 s^{-1}}$, where $\widehat{({\bf A},{\bf B}})$ denotes the angle between the ${\bf A}$ and ${\bf B}$, while $|{\bf A}|$ denotes the standard Euclidean norm of ${\bf A}$. The value of $K_i$ is conventionally chosen to categorise values of $K_f$ above and below the threshold $K_f = 10^{-5} {\rm m^2 s^{-1}}$ as large or small respectively.  

\par 

\begin{figure*}[t]
\noindent\includegraphics[width=39pc,angle=0]{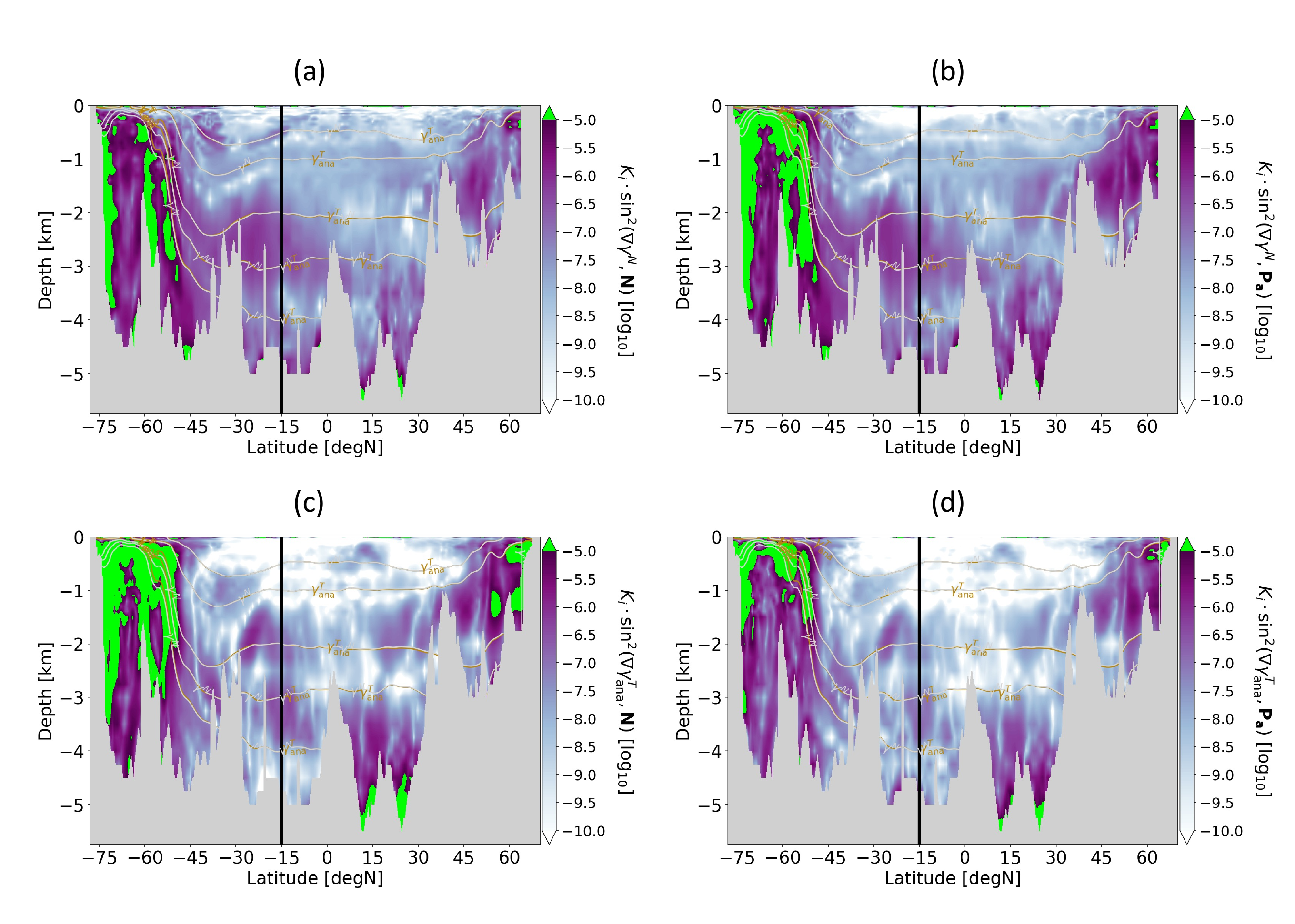}
\\ 
\caption{Latitude/depth section along $30^{\circ}W$ in the Atlantic Ocean of the decimal logarithm of the effective diapycnal mixing like metric quantifying (a) the N-neutrality of $\gamma^n$; (b) the P-neutrality of $\gamma^n$; c) the N-neutrality of $\gamma^T_{analytic}$; d) the P-neutrality of $\gamma^T_{analytic}$.} 
\label{f2} 
\end{figure*}

\begin{figure*}[t]
\centerline{ 
\includegraphics[width=23pc]{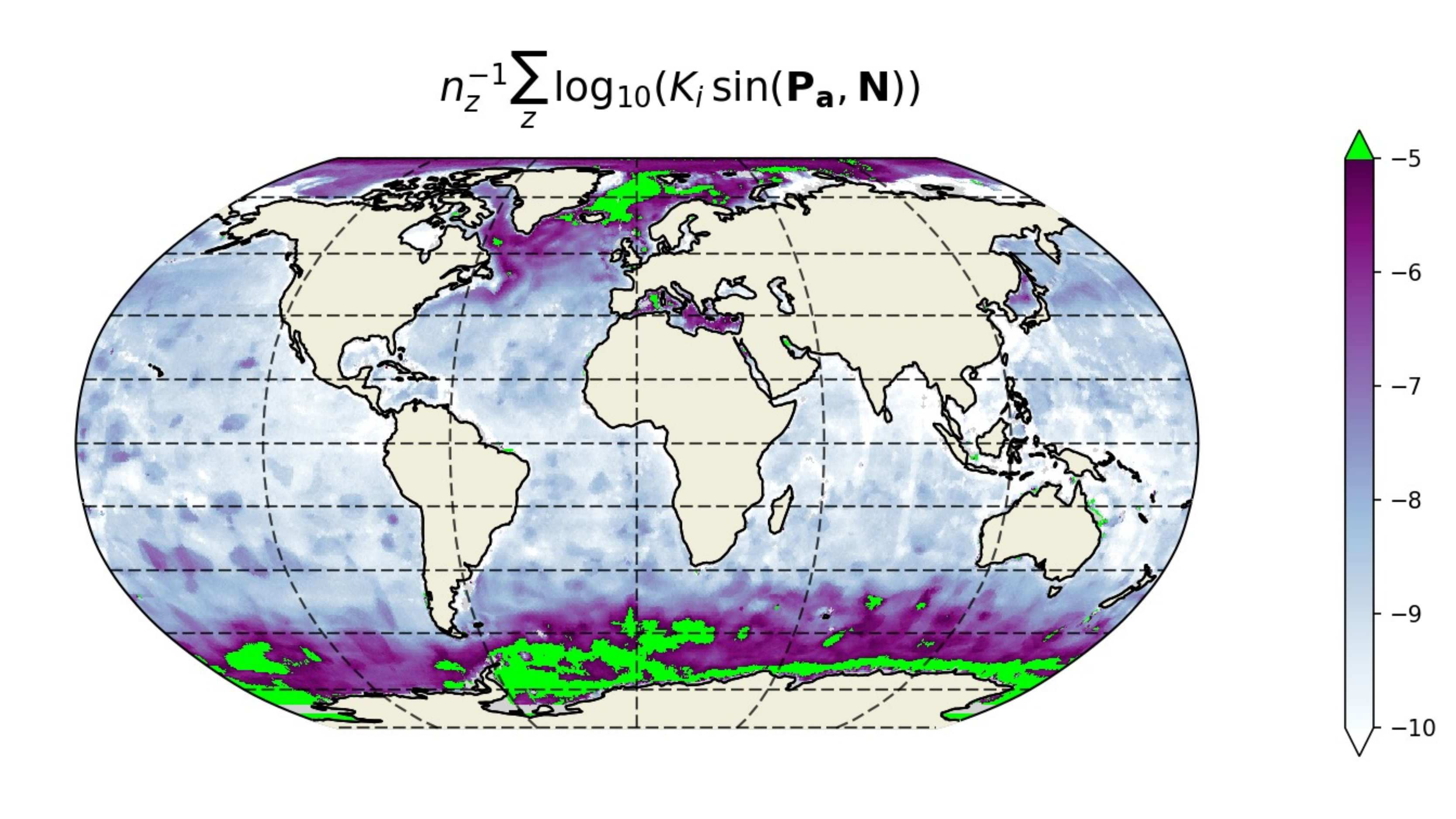}}
\caption{Vertical mean of the decimal logarithm of the effective diapycnal diffusivity like metric illustrating the geographical distribution of the differences between ${\bf N}$ and ${\bf P}_a$ (approximated by ${\bf P}_{a2} $)} 
\label{f3} 
\end{figure*}

Fig. \ref{f1} shows the statistics of $K_f({\bf N},\overline{\bf N}) = K_f ({\bf N},{\bf P}_{a2})$ for the main oceanic basins (left panel) versus for the polar oceans (right panel), with the statistics for whole oceans in the background, which confirm our theoretical prediction that the differences between P-neutrality and N-neutrality should be the largest where fluid parcels are the furthest away from their equilibrium position, that is where $p$ differs the most from $p_r$. This result is further evidenced in Figs. \ref{f2} and \ref{f3} from alternate viewpoints. Interestingly, Fig. \ref{f2} demonstrates that $\gamma^T_{analytic}$ tends to be in general both more P- and N-neutral than $\gamma^n$ outside the polar regions. Overall, Fig. \ref{f3} indicates that thermobaric forces are likely to be important only in the polar regions, but otherwise near negligible in the largest fraction of the oceans, consistent with \citet{Jackett1997}'s statistical analysis of neutral helicity, finding these to be very small in $95\%$ of the oceans. Fig. \ref{f2} (d) also shows that ${\bf P}_{a2}$ is nearly perpendicular to the LRD surfaces outside the polar oceans, thus vindicating the idea that removing the dynamically inactive parts of ${\cal B}_h$ and ${\bf P}_h$ is the key to define a term in the thermodynamic form of the momentum balance indicative of the lateral stirring directions in the oceans. 
\textcolor{black}{The green and dark violet regions in Fig. \ref{f2}(c) and (d) indicate where the lateral stirring directions of $\gamma^T_{analytic}$ are unlikely to be physically realisable and where $\gamma^T_{analytic}$ will need to be corrected in the future. We acknowledge that in those regions, it is possible that $\gamma^n$ (and other ANS perhaps) might be closer to the `true' LMS than the LRD surfaces, at least for the time being. }

\section{Summary and conclusions}
\label{conclusions} 

\textcolor{black}{
In this paper, we have used an energetics approach rooted in \citet{Tailleux2018} local APE theory to develop a first-principles theory of lateral stirring and lateral stirring surfaces that regards lateral stirring as the notional form of stirring that minimally perturb the APE of the oceans. Physically, this is essentially equivalent to how \citet{Sverdrup1942} and oceanographers originally approached lateral stirring (who focused on minimally perturbing the ocean stratification and its potential energy), but quite different from \citet{McDougall1987}'s ambiguous buoyancy-forces based redefinition of the neutral directions. Indeed, one of our main results is that lateral stirring in seawater entails work against both buoyancy and thermobaric forces regardless of the lateral stirring directions considered, so that the directions along which the interchange of fluid parcels do not experience any restoring buoyancy forces that form the basis for \citet{McDougall1987}'s approach do not appear to exist in the oceans. In reality, lateral stirring along the neutral directions can only exist if it is somehow possible for the work against thermobaric forces to be compensated by work against buoyancy forces of opposite sign. Physically, this represents a strong constraint difficult to achieve in reality, which calls into question the physical realisability of isoneutral stirring where thermobaric forces are large, which could perhaps explain, at least partly, why \citet{vanSebille2011} found $\sigma_2$ to outperform $\gamma^n$ and $\sigma_0$ for tracing Labrador Sea Water from its formation regions to the Abaco line in the Gulf Stream area. Our theory also establishes that the actual neutral directions in a continuously stratified binary fluid such as seawater are not the directions perpendicular to the standard N-neutral vector, contrary to what has been assumed so far, but the directions perpendicular to an APE-based form of the P vector previously identified by \citet{Nycander2011}. Physically, the P-neutral directions are those along which stirring minimally perturb the local APE density $\Pi_2$, consistent with our interpretation of \citet{Sverdrup1942}'s original definition of lateral stirring. Importantly, such a result naturally connects the theory of lateral stirring with the theory of diapycnal mixing defining the later in terms of the APE dissipation rate. In fact, where thermobaric forces are large, lateral and vertical stirring appear to be strongly coupled, suggesting that the two processes cannot be studied or understood independently from each other in seawater. In practice, the N-neutral and P-neutral directions are found to approximately coincide where thermobaric forces are weak, that is in most of the oceans except in the polar and Gulf stream regions, which are where the identification of the `right' lateral mixing surfaces appear to be the most challenging theoretically. 
}

\par 

\textcolor{black}{
Where thermobaric forces are small enough to be neglected, our theory establishes that lateral stirring should primarily takes place along the LRD surfaces entering APE theory, as in such regions the LRD surfaces are very accurately neutral and seawater approximately behaves like a simple fluid. Our theory also establishes that in such regions, empirical ANS and LRD surfaces should approximately coincide with each other, which we confirmed empirically by comparing \citet{Tailleux2021} $\gamma^T_{analytic}$ with a new quasi-material approximant $\gamma^n_{material}$ directly in thermohaline space, thus further confirming the previous conclusions of \citet{Tailleux2016b} and \citet{Tailleux2021}. Where thermobaric forces are large, however, lateral stirring along the LRD surfaces entails a non-zero energy cost due to work against thermobaric forces and can no longer occur without interactions with other energy compartments of the system that are expected to cause lateral stirring to occur along different directions. At this stage, our theory remains insufficient to predict what these directions should be, but we are confident that this can be remedied in a future study; for the time being, we acknowledge that $\gamma^n$ or some other ANS might be a better predictor of the `true' lateral mixing surfaces than the LRD surfaces. 
}

\par 

\textcolor{black}{
While our results appear to detract with \citet{McDougall1987}'s view of lateral stirring, they appear to support his view, at least to some extent and for different reasons, that thermobaricity might lead to a form of dianeutral upwelling without a signature in $\varepsilon_k$ \citep{McDougall2003b}. Indeed, Prof. McDougall's explanation for it (as far as we understand) is that this should be viewed as a consequence of the helical character of finite amplitude neutral trajectories. How this is supposed to work and how this could be tested is unclear, however, because the helical behaviour of finite amplitude trajectories is due to the artificial sinks/sources of heat and salt that are necessary to keep such trajectories neutral, as pointed out by \citet{Tailleux2016a}. Indeed, if neutral trajectories conserved their heat and salt content, they would return to exactly the same position they started from in a closed loop. While one may try to justify these artificial sinks/sources of heat and salt as arising from the mixing of the fluid parcel with its environment, it is not clear that such mixing would necessarily be realisable or compatible with down-gradient mixing or the existence of the required sources of energy necessary to sustain it. In any case, the associated diapycnal dispersion would be expected to have a signature in the dissipation of temperature and salinity variance if not in $\varepsilon_k$. In our theory, on the other hand, the possibility for this form of dispersion is seen as a consequence of the coupling between lateral and vertical stirring characterising energetically closed $(\Delta E =0)$ stirring, provided that thermobaric and buoyancy forces are destabilising and stabilising respectively. To the extent that this is possible, this would indicate thermobaric energy as the form of energy sustaining this form of vertical dispersion. The way forward to study it, therefore, will require the development of a theory of thermobaric energy and of thermobaric forces and of their interactions with buoyancy forces. Note that although thermobaricity is central to many hypothesised processes and phenomena \citep{Muller1986,Straub1999,Akimoto1999_part1,Adkins_etal_2005,Su2016part1,Su2016part2,deSzoeke2004,Hallberg2005,Stewart2016}, no comprehensive theory of thermobaric energy and thermobaric forces exist yet that we know of. Interestingly, our theory predicts this form of vertical dispersion to have an infinite dissipation ratio or mixing efficiency $\varepsilon_p/\varepsilon_k = +\infty$, which makes it potentially important for resolving the `missing mixing controversy' \citep{Munk1998}, which we plan on investigating in future work.
}

\par 

\textcolor{black}{
To sum up, we believe that our theory represents a major advance that will enable the rigorous study of lateral stirring and lateral stirring surfaces in terms of mathematically well posed problems issued from the study of the equations of motion, thus allowing oceanographers to finally break away from the two-parcel heuristics and subjective approaches that have been the main basis for the discipline for so long. In terms of immediate future developments, we plan to show in a subsequent study how to use the budgets of resolved and unresolved APE to fix the mixing directions of \citet{Redi1982} rotated diffusion tenors in a physically-based way. Indeed, a key implication of our results is to suggest that neutral rotation tensors \citep{Griffies1998b}, \citet{Shao2020} are potentially inaccurate, and hence that they could be responsible for part of the spurious diapycnal mixing still plaguing most numerical ocean models but generally attributed entirely to numerical mixing, e.g., \citet{Megann2018}. Finally, the fact that APE theory appears central for identifying the right neutral directions in a compressible ocean, via the derivation of the available thermodynamic form of momentum balance, implies that the importance of the concept of APE goes much beyond understanding ocean energetics  \citet{Tailleux2009,Hughes2009,Tailleux2010,Tailleux2010b}, and that it should also play a central role for ocean circulation theory, as we hope to demonstrate in future studies.
}

\clearpage
\acknowledgments
The authors gratefully acknowledge extensive comments by Geoff Stanley, Jacques Vanneste, Jonas Nycander, and an anonymous referee, as well as useful remarks from Thomas Dubos, Guillaume Roullet, Alain Colin de Verdi\`ere, Olivier Arzel, and Peter Rhines, which significantly contributed to improve the manuscript. This research has been supported by the NERC-funded OUTCROP project (grant no. NE/R010536/1).

%
%
\datastatement
The WOCE Global Ocean Climatology 1990-1998 (file '{\tt wghc\_params.nc}') used in this study is available at doi:10.25592/uhhfdm.8987. 
The $\gamma^n$ software was downloaded from 
{\tt https://www.teos\-10.org/preteos10\_software}.  


\appendix[A]
\appendixtitle{Navier-Stokes equations for compressible seawater}
\label{appendixa} 

The Navier-Stokes equations describing the motions of two-component compressible seawater are
\begin{equation}
      \frac{D{\bf v}}{Dt} = - 2{\boldsymbol{\Omega}} \times {\bf v} - \frac{1}{\rho} \nabla p
        - \nabla \Phi + {\bf F} ,
       \label{PGVBF_NSE} 
\end{equation}
\begin{equation}
        \frac{D\rho}{Dt} + \rho \nabla \cdot {\bf v} = 0 ,
        \label{continuity}
\end{equation}
\begin{equation}
        \frac{D\eta}{Dt} = \dot{\eta}, \qquad \frac{DS}{Dt} = \dot{S}, 
        \label{materiality} 
\end{equation}
\begin{equation}
      \upsilon = \upsilon (\eta,S,p) = \frac{\partial h}{\partial p} , 
      \label{nonlinear_eos} 
\end{equation}
where ${\bf v}=(u,v,w)$ is the 3D velocity field, $p$ is pressure, $\rho$ is density, ${\boldsymbol{\Omega}}$ is Earth's rotation vector, ${\bf F}$ is the frictional force. 
$\upsilon = 1/\rho$ is the specific volume, $h = h(\eta,S,p)$ is the specific enthalpy, $\eta$ is the specific entropy, $\Phi(z)=gz$ is the geopotential with $g$ the gravitational acceleration and $z$ height increasing upward.

\appendix[B]
\appendixtitle{Canonical variables and derivatives of $\Pi$, $\Pi_1$ and $\Pi_2$} 

The definition of the local potential energy densities $\Pi$, $\Pi_1$ and $\Pi_2$ involve both thermodynamic $(\eta,S,\rho,p)$ and geometric variables $(z)$. \textcolor{black}{In thermodynamics, the most fundamental set of variables are the canonical (or natural) variables. For a particularly clear and lucid discussion of such variables, see \citet{Alberty1994}. Canonical variables are easily obtained by differentiating everything in sight and examining what is left.}  Thus in the case of $\Pi$
\begin{equation}
    \Pi = h(\eta,S,p) - h(\eta,S,p_r) + g(z-z_r) 
    + \frac{p_0(z)-p}{\rho}  ,
    \label{pi_formula} 
\end{equation}
this approach yields
\begin{equation}
\begin{split}
   {\rm d} \Pi = & 
   T {\rm d} \eta + \mu {\rm d} S + \frac{{\rm d}p}{\rho}
   - T_r {\rm d} \eta - \mu_r {\rm d} S 
    -  \frac{{\rm d}p_r}{\rho_r}\\
   + & g ( {\rm d} z - {\rm d} z_r ) -
  \delta p {\rm d} \upsilon + 
   \frac{{\rm d}(p_0(z)-p)}{\rho}  ,
\end{split} 
   \label{gradient_pi_step1} 
\end{equation}  
where as in the text, the suffix `r' denotes variables estimated at the reference pressure $p_r = p_0(z_r)$,
with $\delta p = p-p_0(z)$. Now, using the fact that $\rho_r = \rho(S,\eta,p_r) = \rho_0(z_r)$ by virtue of $z_r$ satisfying the LNB equation (\ref{LNB_equation}), 
\begin{equation}
     \frac{{\rm d} p_r }{\rho_r} + g {\rm d} z_r 
     = - \frac{\rho_0(z_r)g}{\rho_r} {\rm d} z_r
     + g {\rm d}  z_r = 0 , 
\end{equation}
so that (\ref{gradient_pi_step1}) simplifies to
\begin{equation}
    {\rm d} \Pi = (T-T_r) {\rm d} \eta + 
    (\mu-\mu_r) {\rm d} S 
    -  \delta p {\rm d} \upsilon 
    + g \left ( 1 - \frac{\rho_0(z)}{\rho} \right )
    {\rm d} z  ,
    \label{gradient_pi_step2} 
\end{equation}
where we used the result that ${\rm d} p_0(z) = - \rho_0(z) g {\rm d} z$. 
\textcolor{black}{Eq. (\ref{gradient_pi_step2}) shows that after all simplifications, we are left with terms multiplying the elementary differentials for $(\eta,S,\upsilon,z)$, which hence take as} the canonical variables of $\Pi$. Proceeding similarly with $\Pi_1 = h(\eta,S,p)-h(\eta,S,p_0(z)) + (p_0(z)-p)/\rho$ and $\Pi_2 = h(\eta,S,p_0(z))-h(\eta,S,p_r)+g(z-z_r)$, it is easily verified that
\begin{equation}
{\rm d} \Pi_1 = 
(T-T_b) {\rm d} \eta + (\mu-\mu_b) {\rm d} S 
- \delta p{\rm d} \upsilon + 
\frac{g\rho_0(z)}{\rho_b} 
 \left ( 1 - \frac{\rho_b}{\rho} \right ) {\rm d} z ,
 \label{gradient_pi1} 
\end{equation}
\begin{equation}
    {\rm d} \Pi_2 = 
    (T_b-T_r) {\rm d} \eta + (\mu_b - \mu_r) {\rm d}  S  
    + g\left ( 1 - \frac{\rho_0(z)}{\rho_b} \right ) 
    {\rm d} z ,
    \label{gradient_pi2} 
\end{equation}
where as in the text, the suffix `b' denotes variables estimated at $p_0(z)$. Eqs. (\ref{gradient_pi1}) and (\ref{gradient_pi2}) thus establish that $(\eta,S,\upsilon,z)$ and $(\eta,S,z)$ are the canonical variables of $\Pi_1$ and $\Pi_2$ respectively. It may be verified that summing (\ref{gradient_pi1}) and (\ref{gradient_pi2}) recovers (\ref{gradient_pi_step2}), as expected. 
\textcolor{black}{Eqs. (\ref{gradient_pi1}) and (\ref{gradient_pi2}) motivate the definitions
\begin{equation}
     {\bf P}_{a1} = \left . \frac{\partial \Pi_1}{\partial \eta} 
     \right |_{S,\upsilon,z} \nabla \eta + 
     \left . \frac{\partial \Pi_1}{\partial S} \right |_{\eta,\upsilon,z}
     \nabla S 
      = (T-T_r) \nabla \eta + (\mu-\mu_b) \nabla S , 
      \label{pa1_definition} 
\end{equation}
\begin{equation}
    {\bf P}_{a2} = \left . \frac{\partial \Pi_2}{\partial \eta} 
    \right |_{S,z} \nabla \eta + \left . \frac{\partial \Pi_2}{\partial S} 
    \right |_{\eta,z} \nabla S = (T_b - T_r) \nabla \eta 
    + (\mu_b - \mu_r) \nabla S . 
    \label{pa2_definition} 
\end{equation}
}
Eq. (\ref{gradient_pi1}) implies for the Lagrangian derivative of $\Pi_1$ 
\begin{equation}
    \frac{D\Pi_1}{Dt} = 
    (T-T_b) \frac{D\eta}{Dt}
    + (\mu-\mu_b) \frac{DS}{Dt}
    - \delta p \frac{D\upsilon}{Dt} 
    + g \frac{\rho_0(z)}{\rho_b} 
    \left ( 1 - \frac{\rho_b}{\rho} \right ) w
\end{equation}
By definition, $D\Pi_1/Dt$ may also be written as
\begin{equation} 
\begin{split} 
  \frac{D\Pi_1}{Dt} = & 
  \frac{\partial \Pi_1}{\partial t}
  + {\bf v}\cdot \nabla \Pi_1 \\  
    = & \frac{\partial \Pi_1}{\partial t}
    + {\bf v} \cdot {\bf P}_{a1} 
    - \delta p {\bf v}\cdot \nabla \upsilon 
    + \frac{g\rho_0(z)}{\rho_b} 
    \left ( 1 - \frac{\rho_b}{\rho} \right ) w  
\end{split}
\end{equation} 
Equating the two expressions thus implies
\begin{equation}
    \frac{\partial \Pi_1}{\partial t}
    + \delta p \frac{\partial \upsilon}{\partial t}
    + {\bf v}\cdot {\bf P}_{a1} 
    = (T-T_b) \dot{\eta} + (\mu-\mu_b) \dot{S} ,
\end{equation}
\textcolor{black}{where $\dot{S}$ and $\dot{\eta}$ are shorthand for $DS/Dt$ and $D\eta/Dt$ respectively.} Applying the same idea to $\Pi_2$ yields
\begin{equation}
    \frac{\partial \Pi_2}{\partial t}
    + {\bf v} \cdot {\bf P}_{a2} 
    = (T_b-T_r) \dot{\eta} + (\mu_b - \mu_r) \dot{S} 
\end{equation}
These results are important to relate the work terms ${\bf v}\cdot {\bf P}_{a1}$ and ${\bf v}\cdot {\bf P}_{a2}$ to local Eulerian time derivatives of $\Pi_1$, $\Pi_2$, and $\upsilon$, as well as to irreversible mixing processes.

\appendix[C]
\appendixtitle{Alternative expressions for ${\bf P}_a$}
In practical applications, it is useful to have expressions of ${\bf P}_a$ in terms of the more commonly used in-situ temperature $T$, potential temperature $\theta$, or Conservative Temperature $\Theta$. Using the passage relationships  
\[
    T {\rm d}\eta + \mu {\rm d}S = \frac{T c_{p\theta}}{\theta} {\rm d}\theta 
     + \left ( \mu - T \frac{\partial \mu_{\theta}}{\partial \theta} \right ) {\rm d}S 
\]
\[
     = \frac{T c_{p0}}{\theta} {\rm d}\Theta + 
     \left ( \mu - \frac{T \mu_{\theta}}{\theta} \right ) \,{\rm d}S 
\]
\begin{equation} 
     = c_p \left ( {\rm d}T - \Gamma {\rm d}p \right ) 
     + \left ( \mu - T \frac{\partial \mu}{\partial T} \right ) {\rm d}S ,
     \label{passage_relations} 
\end{equation}
e.g., \citet{Tailleux2010,Tailleux2015b}, yields
\begin{equation} 
\begin{split}
  {\bf P}_a = & (T-T_r) \nabla \eta + (\mu-\mu_r) \nabla S \\
        = & \frac{T-T_r}{T} c_p \left ( \nabla T - \Gamma \nabla p
  \right ) + \left ( \mu - \mu_r - (T-T_r) \frac{\partial \mu}{\partial T} 
  \right ) \nabla S \\
   = & 
   \left ( \frac{T-T_r}{\theta} \right ) c_{p\theta} \nabla \theta 
   + \left ( \mu - \mu_r - (T-T_r) \frac{\partial \mu_{\theta}}{\partial \theta} 
   \right ) \nabla S \\
     = & \left ( \frac{T-T_r}{\theta} \right ) c_{p0} \nabla \Theta 
     + \left ( \mu - \mu_r - \left ( \frac{T-T_r}{\theta} \right ) \mu_{\theta} \right ) \nabla S ,
\end{split} 
     \label{buoyancy_force_expressions} 
\end{equation}
where $\Gamma = 
\alpha T/(\rho c_p)$ is the adiabatic lapse rate,
$c_{p\theta} = c_p(\eta,S,0)$, $\mu_{\theta} = \mu(\eta,S,0)$, while $c_{p0}$ is the constant reference specific heat capacity underlying TEOS-10. 

\bibliographystyle{ametsoc2014}
\bibliography{energetics}

\begin{thebibliography}{74}
\providecommand{\natexlab}[1]{#1}
\providecommand{\url}[1]{\texttt{#1}}
\renewcommand{\UrlFont}{\rmfamily}
\providecommand{\urlprefix}{URL }
\expandafter\ifx\csname urlstyle\endcsname\relax
  \providecommand{\doi}[1]{doi:\discretionary{}{}{}#1}\else
  \providecommand{\doi}{doi:\discretionary{}{}{}\begingroup
  \urlstyle{rm}\Url}\fi
\providecommand{\eprint}[2][]{\url{#2}}

\bibitem[{Adkins et~al.(2005)Adkins, Ingersoll,, and
  Pasquero}]{Adkins_etal_2005}
Adkins, J.~F., A.~P. Ingersoll, and C.~Pasquero, 2005: Rapid climate change and
  conditional instability of the glacial deep ocean from the thermobaric effect
  and geothermal heating. \textit{Quat. Science Rev.}, \textbf{24}, 581--594.

\bibitem[{Akimoto(1999)}]{Akimoto1999_part1}
Akimoto, K., 1999: Open-ocean deep convection due to thermobaricity 1. scaling
  argument. \textit{J. Geophys. Res.}, \textbf{104}, 5225--5234.

\bibitem[{Alberty(1994)}]{Alberty1994}
Alberty, R.~A., 1994: Legendre transforms in chemical thermodynamics.
  \textit{Chemical Reviews}, \textbf{94~(6)}, 1457--1482,
  \doi{10.1021/cr00030a001}.

\bibitem[{Boning et~al.(1995)Boning, Holland, Bryan, Danabasoglu,, and
  McWilliams}]{Boning1995}
Boning, C., W.~R. Holland, F.~O. Bryan, G.~Danabasoglu, and J.~C. McWilliams,
  1995: An overlooked problem in model simulations o fthe thermohaline
  circulation and heat transport in the atlantic ocean. \textit{J. Climate},
  \textbf{8}, 515--523.

\bibitem[{Crocco(1937)}]{Crocco1937}
Crocco, L., 1937: Eine neue {S}tromfunktion f\"{u}r die {E}rforschung der
  {B}ewegung der {G}ase mit {R}otation. \textit{ZAMM - Zeitschrift f\"{u}r
  Angewandte Mathematik und Mechanik}, \textbf{17}, 1--7,
  \doi{10.1002/zamm.19370170103}.

\bibitem[{de~Szoeke(2000)}]{deSzoeke2000a}
de~Szoeke, R.~A., 2000: Equations of motion using thermodynamic coordinates.
  \textit{J. Phys. Oceanogr.}, \textbf{30}, 2184--2829.

\bibitem[{de~Szoeke(2004)}]{deSzoeke2004}
de~Szoeke, R.~A., 2004: An effect of the thermobaric nonlinearity of the
  equation of state: a mechanism for sustaining solitary {R}ossby waves.
  \textit{J. Phys. Oceanogr.}, \textbf{34}, 2042--2056.

\bibitem[{de~Szoeke and Springer(2000)de~Szoeke, and Springer}]{deSzoeke2000b}
de~Szoeke, R.~A., and S.~R. Springer, 2000: Orthobaric density: A thermodynamic
  variable for ocean circulation studies. \textit{J. Phys. Oceanogr.},
  \textbf{30}, 2830--2852.

\bibitem[{de~Szoeke and Springer(2009)de~Szoeke, and Springer}]{deSzoeke2009}
de~Szoeke, R.~A., and S.~R. Springer, 2009: The materiality and neutrality of
  neutral density and orthobaric density. \textit{J. Phys. Oceanogr.},
  \textbf{39}, 1779--1799.

\bibitem[{Dewar and McWilliams(2019)Dewar, and McWilliams}]{Dewar2019}
Dewar, W.~K., and J.~C. McWilliams, 2019: On energy and turbulent mixing in the
  thermocline. \textit{J. Adv. Modeling Earth System}, \textbf{11}, 578--596,
  \doi{Earth Systems, 11, 578–596. 10.1029/2018MS001502}.

\bibitem[{Eden(2015)}]{Eden2015}
Eden, C., 2015: Revisiting the energetics of the ocean in boussinesq
  approximation. \textit{J. Phys. Oceanogr.}, \textbf{45}, 630--637.

\bibitem[{Eden and Willebrand(1999)Eden, and Willebrand}]{Eden1999}
Eden, C., and J.~Willebrand, 1999: Neutral density revisited. \textit{Deep Sea
  Res. Part II}, \textbf{46}, 33--54.

\bibitem[{Fofonoff(1998)}]{Fofonoff1998}
Fofonoff, N.~P., 1998: Nonlinear limits to ocean thermal structure. \textit{J.
  Mar. Res.}, \textbf{56}, 793--811, \doi{10.1357/002224098321667378}.

\bibitem[{Foster and Carmack(1976)Foster, and Carmack}]{Foster1976}
Foster, T.~D., and E.~C. Carmack, 1976: Frontal zone mixing and {A}ntarctic
  {B}ottom {W}ater formation in the southern {W}eddell {S}ea. \textit{Deep Sea
  Res.}, \textbf{23}, 301--317.

\bibitem[{Gargett and Holloway(1984)Gargett, and Holloway}]{Gargett1984b}
Gargett, A.~E., and G.~Holloway, 1984: Dissipation and diffusion by internal
  wave breaking. \textit{J. Mar. Res.}, \textbf{42}, 15--27.

\bibitem[{Gouretski and Koltermann(2004)Gouretski, and
  Koltermann}]{Gouretski2004}
Gouretski, V.~V., and K.~P. Koltermann, 2004: {WOCE} global hydrographic
  climatology. \textit{Berichte des Bundesamtes f\"{u}r Seeshifffahrt und
  Hydrographie Tech. Rep. 35/2004}, 49pp.

\bibitem[{Gregg(2021)}]{Gregg2021}
Gregg, M.~C., 2021: \textit{Ocean Mixing}. Cambridge University Press.

\bibitem[{Griffies et~al.(1998)Griffies, Gnanadesikan, Pacanowski, Larichev,
  Dukowicz,, and Smith}]{Griffies1998b}
Griffies, S.~M., A.~Gnanadesikan, R.~C. Pacanowski, V.~D. Larichev, J.~K.
  Dukowicz, and R.~D. Smith, 1998: Isoneutral diffusion in a z-coordinate ocean
  model. \textit{J. Phys. Oceanogr.}, \textbf{28}, 805--830.

\bibitem[{Hallberg(2005)}]{Hallberg2005}
Hallberg, R., 2005: A thermobaric instability of lagrangian vertical coordinate
  ocean models. \textit{Ocean Modelling}, \textbf{8}, 279--300.

\bibitem[{Harris et~al.(2022)Harris, Tailleux, Holloway,, and
  Vidale}]{Harris2022}
Harris, B.~L., R.~Tailleux, C.~E. Holloway, and P.~L. Vidale, 2022: A moist
  available potential energy budget for an axisymmetric tropical cyclone.
  \textit{J. Atmos. Sci.}, \textbf{79}, 2493--2513,
  \doi{10.1175/JAS-D-22-0040.1}.

\bibitem[{Hochet et~al.(2019)Hochet, Tailleux, Ferreira,, and
  Kuhlbrodt}]{Hochet2019}
Hochet, A., R.~Tailleux, D.~Ferreira, and T.~Kuhlbrodt, 2019: Isoneutral
  control of effective diapycnal mixing in numerical ocean models with neutral
  rotated diffusion tensors. \textit{Ocean Science}, \textbf{15}, 21--32.

\bibitem[{Huang(2005)}]{Huang2005}
Huang, R.~X., 2005: Available potential energy in the world's oceans.
  \textit{J. Mar. Res.}, \textbf{63}, 141--158.

\bibitem[{Hughes et~al.(2009)Hughes, Hogg,, and Griffiths}]{Hughes2009}
Hughes, G.~O., A.~M. Hogg, and R.~W. Griffiths, 2009: Available potential
  energy and irreversible mixing in the meridional overturning circulation.
  \textit{J. Phys. Oceanogr.}, \textbf{39}, 3130--3146.

\bibitem[{Iselin(1939)}]{Iselin1939}
Iselin, C.~O., 1939: The influence of vertical and lateral turbulence on the
  characteristics of the waters at mid-depth. \textit{Trans. Amer. Geophys.
  Union}, \textbf{20}, 414--417.

\bibitem[{Jackett and McDougall(1997)Jackett, and McDougall}]{Jackett1997}
Jackett, D.~R., and T.~J. McDougall, 1997: A neutral density variable for the
  world's oceans. \textit{J. Phys. Oceanogr.}, \textbf{27}, 237--263.

\bibitem[{Klocker et~al.(2009)Klocker, McDougall,, and Jackett}]{Klocker2009}
Klocker, A., T.~J. McDougall, and D.~R. Jackett, 2009: A new method for forming
  approximately neutral surfaces. \textit{Ocean Sci.}, \textbf{5}, 155--172.

\bibitem[{Lang et~al.(2020)Lang, Stanley, McDougall,, and Barker}]{Lang2020}
Lang, Y., G.~J. Stanley, T.~J. McDougall, and P.~M. Barker, 2020: A
  pressure-invarient neutral density variable for the world's oceans.
  \textit{J. Phys. Oceanogr.}, 3585--3604.

\bibitem[{Lindborg and Brethouwer(2008)Lindborg, and Brethouwer}]{Lindborg2008}
Lindborg, E., and G.~Brethouwer, 2008: Vertical dispersion by stratified
  turbulence. \textit{J. Fluid Mech.}, \textbf{614}, 303--314.

\bibitem[{McDougall(1987{\natexlab{a}})}]{McDougall1987}
McDougall, T.~J., 1987{\natexlab{a}}: Neutral surfaces. \textit{J. Phys.
  Oceanogr.}, \textbf{17}, 1950--1964.

\bibitem[{McDougall(1987{\natexlab{b}})}]{McDougall1987c}
McDougall, T.~J., 1987{\natexlab{b}}: Thermobaricity, cabbeling, and water
  mass-conversion. \textit{J. Geophys. Res. Oceans}, 5448--5464.

\bibitem[{McDougall(2003)}]{McDougall2003b}
McDougall, T.~J., 2003: Dianeutral upwelling without dissipation of kinetic
  energy. \textit{Near boundary processes and their parameterization.
  Proceedings of the 13th 'Aha Huliko' a Hawaiian Winter Workshop}, P.~Muller,
  and D.~Anderson, Eds., University of Hawaii, Manoa., 229--237.

\bibitem[{McDougall et~al.(2017)McDougall, Groeskamp,, and
  Griffies}]{McDougall2017}
McDougall, T.~J., S.~Groeskamp, and S.~M. Griffies, 2017: Comments on tailleux,
  r. neutrality versus materiality: A thermodynamic theory of neutral surfaces.
  fluids 2016, 1, 32. \textit{Fluids}, \textbf{2~(19)}.

\bibitem[{McDougall and Jackett(1988)McDougall, and Jackett}]{McDougall1988b}
McDougall, T.~J., and D.~R. Jackett, 1988: On the helical nature of neutral
  trajectories in the ocean. \textit{Prog. Oceanogr.}, \textbf{20}, 153--183.

\bibitem[{McDougall and Jackett(2005)McDougall, and Jackett}]{McDougall2005b}
McDougall, T.~J., and D.~R. Jackett, 2005: The material derivative of neutral
  density. \textit{J. Mar. Res.}, \textbf{63}, 159--185.

\bibitem[{Megann(2018)}]{Megann2018}
Megann, A., 2018: Estimating the numerical diapycnal mixing in an
  eddy-permitting ocean model. \textit{Ocean Modell.}, \textbf{121}, 19--33,
  \doi{10.1016/j.ocemod.2017.11.001}.

\bibitem[{Montgomery(1937)}]{Montgomery1937}
Montgomery, R.~B., 1937: A suggested method for representing gradient flow in
  isentropic surfaces. \textit{Bull. Amer. Meteor. Soc.}, \textbf{18},
  210--212.

\bibitem[{Montgomery(1938)}]{Montgomery1938}
Montgomery, R.~B., 1938: Circulation in the upper layers of the southern north
  atlantic, deduced with the use of isentropic analysis. \textit{Pap. Phys.
  Oceanogr. Meteor}, \textbf{6(2)}, 55 pp.

\bibitem[{Muller and Willebrand(1986)Muller, and Willebrand}]{Muller1986}
Muller, P., and J.~Willebrand, 1986: Compressibility effects in the
  thermohaline circulation: a manifestation of temperature-salinity mode.
  \textit{Deep Sea Res. Part A. Oceanographic Research Papers}, \textbf{33},
  559--571.

\bibitem[{Munk and Wunsch(1998)Munk, and Wunsch}]{Munk1998}
Munk, W.~H., and C.~Wunsch, 1998: Abyssal recipes ii: Energetics of tidal and
  wind mixing. \textit{Deep-Sea Res.}, \textbf{45A}, 1977--2010.

\bibitem[{Novak and Tailleux(2017)Novak, and Tailleux}]{Novak2017}
Novak, L., and R.~Tailleux, 2017: On the local view of atmospheric available
  potential energy. \textit{J. Atmos. Sci. subjudice},
  \textbf{http://arxiv.org/abs/1711.08660}.

\bibitem[{Nycander(2011)}]{Nycander2011}
Nycander, J., 2011: Energy conversion, mixing energy, and neutral surfaces with
  a nonlinear equation of state. \textit{J. Phys. Oceanogr.}, \textbf{41},
  28--41.

\bibitem[{Oakey(1982)}]{Oakey1982}
Oakey, N.~S., 1982: Determination of the rate of dissipation of turbulent
  energy from simultaneous temperature and velocity shear microstructure
  measurements. \textit{J. Phys. Oceanogr.}, \textbf{12}, 256--271.

\bibitem[{Pingree(1972)}]{Pingree1972}
Pingree, R.~D., 1972: Mixing in the deep stratified ocean. \textit{Deep Sea
  Res.}, \textbf{19}, 549--561.

\bibitem[{Redi(1982)}]{Redi1982}
Redi, M.~H., 1982: Oceanic isopycnal mixing by coordinate rotation. \textit{J.
  Phys. Oceanogr.}, \textbf{12}, 1154--1158.

\bibitem[{Reid and Lynn(1971)Reid, and Lynn}]{Reid1971}
Reid, J.~L., and R.~J. Lynn, 1971: On the influence of the norwegian-greenland
  and weddell seas upon the bottom waters of the indian and pacific oceans.
  \textit{Deep-Sea Res.}, \textbf{18}, 1063--1088.

\bibitem[{Reid et~al.(1981)Reid, Elliott,, and Olson}]{Reid1981}
Reid, R.~O., B.~A. Elliott, and D.~B. Olson, 1981: Available potential energy:
  a clarification. \textit{J. Phys. Oceanogr.}, \textbf{11}, 5--29,
  \doi{10.1175/1520-0485(1981)011%3C0015:APEAC%3E2.0.CO;2}.

\bibitem[{Saenz et~al.(2015)Saenz, Tailleux, Butler, Hughes,, and
  Oliver}]{Saenz2015}
Saenz, J.~A., R.~Tailleux, E.~D. Butler, G.~O. Hughes, and K.~I.~C. Oliver,
  2015: Estimating lorenz's reference state in an ocean with a nonlinear
  equation of state for seawater. \textit{J. Phys. Oceanogr.}, \textbf{45},
  1242--1257, \doi{10.1175/JPO-D-14-0105.1}.

\bibitem[{Shao et~al.(2020)Shao, Adcroft, Hallberg,, and Griffies}]{Shao2020}
Shao, A.~E., A.~Adcroft, R.~Hallberg, and S.~M. Griffies, 2020: A
  general-coordinate, nonlocal neutral diffusion operator. \textit{J. Adv.
  Modeling Earth System}, \textbf{~(e2019MS001992)},
  \doi{https://doi.org/10.1029/2019MS001992}.

\bibitem[{Stanley(2019{\natexlab{a}})}]{Stanley2019b}
Stanley, G.~J., 2019{\natexlab{a}}: The exact geostrophic streamfunction for
  neutral surfaces. \textit{Ocean Modelling}, \textbf{138}, 107--121.

\bibitem[{Stanley(2019{\natexlab{b}})}]{Stanley2019a}
Stanley, G.~J., 2019{\natexlab{b}}: Neutral surface topology. \textit{Ocean
  Modelling}, \textbf{138}, 88--106.

\bibitem[{Stanley et~al.(2021)Stanley, McDougall,, and Barker}]{Stanley2021}
Stanley, G.~J., T.~J. McDougall, and P.~M. Barker, 2021: Algorithmic
  improvements to finding approximately neutral surfaces. \textit{Journal of
  Advances in Modeling Earth Systems}, \textbf{13~(e2020MS002436)}.

\bibitem[{Stewart and Haine(2016)Stewart, and Haine}]{Stewart2016}
Stewart, K.~D., and T.~W.~N. Haine, 2016: Thermobaricity in the transition
  zones between alpha and beta oceans. \textit{J. Phys. Oceanogr.},
  \textbf{46}, 1805--1821, \doi{10.1175/JPO-D-16-0017.1}.

\bibitem[{Straub(1999)}]{Straub1999}
Straub, D.~N., 1999: On thermobaric production of potential vorticity in the
  ocean. \textit{Tellus}, \textbf{51}, 314--325.

\bibitem[{Su et~al.(2016{\natexlab{a}})Su, Ingersoll, Stewart,, and
  Thompson}]{Su2016part1}
Su, Z., A.~P. Ingersoll, A.~L. Stewart, and A.~F. Thompson, 2016{\natexlab{a}}:
  Ocean convective available potential energy. part i: Concept and calculation.
  \textit{J. Phys. Oceanogr.}, \textbf{46}, 1081--1096.

\bibitem[{Su et~al.(2016{\natexlab{b}})Su, Ingersoll, Stewart,, and
  Thompson}]{Su2016part2}
Su, Z., A.~P. Ingersoll, A.~L. Stewart, and A.~F. Thompson, 2016{\natexlab{b}}:
  Ocean convective available potential energy. part ii: Energetics of
  thermobaric convection and thermobaric cabelling. \textit{J. Phys.
  Oceanogr.}, \textbf{46}, 1097--1115.

\bibitem[{Sverdrup et~al.(1942)Sverdrup, Johnson,, and Fleming}]{Sverdrup1942}
Sverdrup, H., M.~W. Johnson, and R.~H. Fleming, 1942: \textit{Significance of
  $\sigma_t$ surfaces}, 414--416. Prentice-Hall, Inc. New York,
  \urlprefix\url{http://ark.cdlib.org/ark:/13030/kt167nb66r/}.

\bibitem[{Tailleux(2009)}]{Tailleux2009}
Tailleux, R., 2009: On the energetics of turbulent stratified mixing,
  irreversible thermodynamics, boussinesq models, and the ocean heat engine
  controversy. \textit{J. Fluid Mech.}, \textbf{638}, 339--382.

\bibitem[{Tailleux(2010)}]{Tailleux2010}
Tailleux, R., 2010: Entropy versus {APE} production: on the buoyancy power
  input in the oceans energy cycle. \textit{Geophis. Res. Lett.},
  \textbf{37~(L22603)}.

\bibitem[{Tailleux(2013)}]{Tailleux2013b}
Tailleux, R., 2013: Available potential energy density for a multicomponent
  boussinesq fluid with a nonlinear equation of state. \textit{J. Fluid Mech.},
  \textbf{735}, 499--518, \doi{10.1017/jfm.2013.509}.

\bibitem[{Tailleux(2015{\natexlab{a}})}]{Tailleux2015b}
Tailleux, R., 2015{\natexlab{a}}: Observational and energetic constraints on
  the non-conservation of potential/conservative temperature and implications
  for ocean modelling. \textit{Ocean Modell.}, \textbf{88}, 26--37.

\bibitem[{Tailleux(2015{\natexlab{b}})}]{Tailleux2015}
Tailleux, R., 2015{\natexlab{b}}: On the validity of single-parcel energetics
  to assess the importance of internal energy and compressibility effects in
  stratified fluids. \textit{J. Fluid Mech.}, \textbf{767}, R2.

\bibitem[{Tailleux(2016{\natexlab{a}})}]{Tailleux2016b}
Tailleux, R., 2016{\natexlab{a}}: Generalized patched potential density and
  thermodynamic neutral density: Two new physically based quasi-neutral density
  variables for ocean water masses analyses and circulation studies. \textit{J.
  Phys. Oceanogr.}, \textbf{46}, 3571--3584, \doi{10.1175/JPO-D-16-0072.1}.

\bibitem[{Tailleux(2016{\natexlab{b}})}]{Tailleux2016a}
Tailleux, R., 2016{\natexlab{b}}: Neutrality versus materiality: a
  thermodynamic theory of neutral surfaces. \textit{Fluids}, \textbf{1~(32)},
  \doi{10.3390/fluids1040032}.

\bibitem[{Tailleux(2017)}]{Tailleux2017}
Tailleux, R., 2017: Reply to``comment on tailleux, r. neutrality versus
  materiality: a thermodynamic theoryof neutral surfaces. fluids 2016, 1,
  32.''. \textit{Fluids}, \textbf{2~(20)}.

\bibitem[{Tailleux(2018)}]{Tailleux2018}
Tailleux, R., 2018: Local available energetics of multicomponent compressible
  stratified fluids. \textit{J. Fluid Mech.}, \textbf{842~(R1)},
  \doi{10.1017/jfm.2018.196}.

\bibitem[{Tailleux(2021)}]{Tailleux2021}
Tailleux, R., 2021: Spiciness theory revisited, with new views on neutral
  density, orthogonality and passiveness. \textit{Ocean Science}, \textbf{17},
  203--219, \doi{10.5194/os-17-203-2021}.

\bibitem[{Tailleux and Rouleau(2010)Tailleux, and Rouleau}]{Tailleux2010b}
Tailleux, R., and L.~Rouleau, 2010: The effect of mechanical stirring on
  horizontal convection. \textit{Tellus}, \textbf{62A}, 138--153.

\bibitem[{Taylor et~al.(2019)Taylor, de~Bruyn~Kops, Caulfied,, and
  Linden}]{Taylor2019}
Taylor, J.~R., S.~M. de~Bruyn~Kops, C.~P. Caulfied, and P.~F. Linden, 2019:
  Testing the assumptions underlying ocean mixing methodologies using direct
  numerical simulations. \textit{J. Phys. Oceanogr.}, \textbf{49}, 2761--2779,
  \doi{10.1175/JPO-D-19-0033.1}.

\bibitem[{van Sebille et~al.(2011)van Sebille, Baringer, Johns, Meinen, Beal,
  de~Jong,, and van Haken}]{vanSebille2011}
van Sebille, E., M.~O. Baringer, W.~E. Johns, C.~S. Meinen, L.~M. Beal,
  F.~de~Jong, and H.~M. van Haken, 2011: Propagation pathways of classical
  labrador seawater from its source region to $26^{\circ}n$. \textit{J.
  Geophys. Res.}, \textbf{116}, C1207.

\bibitem[{Vazsonyi(1945)}]{Vazsonyi1945}
Vazsonyi, A., 1945: On rotational gas flows. \textit{Quart. Applied Math.},
  \textbf{3}, 29--37.

\bibitem[{Veronis(1975)}]{Veronis1975}
Veronis, G., 1975: The role of models in tracer studies. \textit{Numerical
  models of Ocean Circulation}, \textbf{National Academy of Science}, 133--146.

\bibitem[{Wexler and Montgomery(1941)Wexler, and Montgomery}]{Wexler1941}
Wexler, H., and R.~B. Montgomery, 1941: "stream function" or "acceleration
  potential"? \textit{Bull. Amer. Meteor. Soc.}, \textbf{22}, 44--46,
  \urlprefix\url{https://www.jstor.org/stable/26256733}.

\bibitem[{Winters and d'Asaro(1996)Winters, and d'Asaro}]{Winters1996}
Winters, K.~B., and E.~A. d'Asaro, 1996: Diascalar flux and the rate of fluid
  mixing. \textit{J. Fluid Mech.}, \textbf{317}, 179--193.

\bibitem[{Winters et~al.(1995)Winters, Lombard, Riley,, and
  d'Asaro}]{Winters1995}
Winters, K.~B., P.~N. Lombard, J.~J. Riley, and E.~A. d'Asaro, 1995: Available
  potential energy and mixing in density stratified fluids. \textit{J. Fluid
  Mech.}, \textbf{289}, 115--128.

\end{thebibliography}

\end{document}